\documentclass[preprint,3p,12pt]{elsarticle}
\usepackage{graphicx} 
\usepackage{caption}
\usepackage{subfigure}
\usepackage{amsmath,amsfonts,amssymb}
\usepackage{algorithm}
\usepackage{algpseudocode}

\newcommand{\Rmnum}[1]{\uppercase\expandafter{\romannumeral #1}}  

\DeclareCaptionLabelFormat{ContinuedFmt}{#1~#2}
\begin{document}
\title{A well-balanced weakly compressible SPH formulation for free-surface flows and its GPU implementation}

\author[HKUST1]{Jiawang Zhang}
\ead{jzhangiw@connect.ust.hk}

\author[HKUST1]{Fengxiang Zhao}
\ead{fzhaoac@connect.ust.hk}

\author[HKUST1,HKUST2]{Jianping Gan}
\ead{magan@ust.hk}

\author[HKUST1,HKUST2,HKUST3]{Kun Xu\corref{cor1}}
\ead{makxu@ust.hk}

\address[HKUST1]{Department of Mathematics, Hong Kong University of Science and Technology, Clear Water Bay, Kowloon, Hong Kong}
\address[HKUST2]{Center for Ocean Research in Hong Kong and Macau (CORE), Hong Kong University of Science and Technology, Clear Water Bay, Kowloon, Hong Kong}
\address[HKUST3]{Shenzhen Research Institute, Hong Kong University of Science and Technology, Shenzhen, China}
\cortext[cor1]{Corresponding author}

\begin{abstract}
    This study proposes a well-balanced formulation of weakly compressible smoothed particle hydrodynamics (WCSPH) for free-surface flows, which preserves hydrostatic equilibrium exactly at the discrete level—a property essential for reliable long-term simulations. Although well-balanced schemes are well established for mesh-based methods, the property remains largely unaddressed in WCSPH, where the particle approximation of the pressure gradient fails to balance the gravitational force exactly.
    The imbalance stems from two difficulties: the nonlinearity of the pressure-gradient-over-density term, and the approximation error of gradients evaluated by particle summation. The first is resolved by introducing an auxiliary potential variable that recasts the nonlinear term as the gradient of a single scalar, which reduces to a linear function of position under hydrostatic conditions. The second is resolved by a Riemann-based gradient approximation with kernel correction, which is first-order consistent and recovers linear fields exactly. These two ingredients ensure that the discrete potential gradient balances gravitational force exactly. Widely used techniques, including $\delta-$SPH, particle shifting and tensile instability control, are readily incorporated. The formulation is further extended to three dimensions and implemented on GPU with architecture-tailored optimizations. Hydrostatic tests with rectangular, triangular and Gaussian bottom topographies show that the proposed formulation attains the well-balanced property to machine precision, reducing the spurious velocity error of conventional SPH from $10^{-3}$ to the order of $10^{-13}$. More complex benchmarks confirm its robustness, accuracy and low pressure oscillation, with simulations of up to 17.53 million particles performed on a single consumer-grade GPU.
\end{abstract}
\begin{keyword}
    Smoothed particle hydrodynamics, well-balanced formulation, GPU-acceleration, free surface flows 
\end{keyword}
\maketitle

\section{Introduction}
Free-surface flows, which involve either a free liquid surface or interfaces separating immiscible fluids, are ubiquitous and critically important in numerous natural phenomena \cite{waveBreak} and  engineering applications, attracting widespread attention from both the scientific and industrial communities \cite{multiphase}. Nevertheless, the long-term evolution and complex dynamics of moving interfaces continue to pose significant challenges for the accurate and robust simulation of free-surface flows.

Most existing numerical methods for free-surface flows are based on mesh-based frameworks combined with interface-capturing techniques, such as the Volume-of-Fluid (VOF) method \cite{vof}, the Level Set (LS) method \cite{ls}, and their variants \cite{clsvof}. The inherent mesh entanglement problem and numerical dissipation make mesh-based methods difficult for long-term free surface flows with large deformations. In view of these limitations, mesh-free particle-based methods have received increasing attention, with SPH being a representative example \cite{sph1,sph2}, where the computational domain is represented by a set of arbitrarily distributed particles. As the particles move, the free surface is naturally identified from a subset of particles
without any additional interface-tracking procedures, which makes such methods suitable for flows with large free surface deformation. With the development of several key numerical techniques, including $\delta-$SPH \cite{delta-sph, delta-sph2}, particle shifting techniques \cite{pst}, and tensile instability control \cite{tic}, SPH has become a robust and widely-used approach for free surface flows, particularly with violent deformation \cite{sph-frsurf}. Despite these substantial progresses  \cite{sph-review1,sph-review2}, several difficulties remain to be resolved before SPH can be applied to long-term, high-fidelity simulations, among which consistency and the well-balanced property are prominent.

The well-balanced property is of vital importance for long-term simulations of nearly hydrostatic states, such as the long-distance propagation of tsunamis over the deep ocean \cite{tsunami} and long-period, low-amplitude sloshing \cite{longTermSPH}. A numerical scheme is said to be well-balanced if it preserves the discrete hydrostatic equilibrium exactly, so that no spurious motion is generated from a quiescent initial state. Well-balanced numerical methods have been thoroughly investigated within mesh-based frameworks, particularly for the shallow water equations \cite{swe-gks1,swe-gks2}, and several well-balanced SPH formulations have likewise been developed for shallow water flows \cite{wb-pp-sph-swe,wb-sph-ale-swe}. In contrast, the well-balanced property of weakly compressible SPH for the Navier–Stokes (NS) equations has not yet been fully studied or resolved. In conventional weakly compressible SPH, the widely used symmetric approximation of the pressure gradient fails to balance the volumetric gravitational force exactly, generating spurious flows that  accumulate over long-term evolution and eventually contaminate the bulk flow or even displace the free surface \cite{longTermSPH}. Constructing a well-balanced SPH formulation for the Navier–Stokes equations involves two principal difficulties. First, the nonlinear pressure-gradient-over-density term in NS equations must balance the gravitational force exactly in order to satisfy the hydrostatic condition. Second, the hydrostatic solution of the density and pressure follows an exponential distribution under the weakly compressible equation of state, so that a particle approximation capable of reproducing an exponential function is required to discretize the pressure gradient and recover the hydrostatic equilibrium which seems intractable particularly for disordered particle distributions. These two aspects together constitute the core challenge addressed in this work.

A further obstacle to the practical applications of SPH is its high computational cost, particularly for real-world engineering problems that typically require very large numbers of particles, so that efficient parallel implementations are essential for such simulations \cite{gpu-early2}. Although considerable progress has been made in distributed-memory, MPI-based parallelization of SPH on CPU platforms \cite{cpu-wavebreak,cpu-mpi,cpu-mpi2}, the transient motion of the particle set and the intensive interactions among large numbers of neighboring particles make SPH particularly well suited to GPU architectures, particularly in light of the rapid growth of GPU computing power over the past decades \cite{power-sum}. Early GPU implementations of SPH already achieved speedups of two orders of magnitude relative to a single CPU core \cite{gpu-cuda,gpu-splitMegre,gpu-early1}, and GPU-accelerated SPH has since been extended to multiphase flows \cite{gpu-multiphase}, fluid–structure interaction \cite{gpu-fsi}, and incompressible SPH \cite{gpu-isph}, among other applications. These developments highlight the acceleration potential of GPU computing for SPH simulations and motivate the GPU implementation of the well-balanced formulation proposed in the present study.

This study proposes a well-balanced SPH (WBSPH) formulation for the Navier–Stokes equations and extends it to three-dimensional problems accelerated on GPU architectures. The two difficulties identified above are addressed in turn. First, a new form of governing equations is derived by introducing an auxiliary potential variable, through which the nonlinear pressure-gradient-over-density term is recast as the gradient of a single scalar that reduces to a linear function of position under hydrostatic conditions. Second, a Riemann-based particle approximation of the potential gradient with first-order consistency is developed, which reproduces linear fields exactly and therefore balances a constant gravitational force exactly in the hydrostatic equilibrium state. Widely used stabilization techniques, including $\delta-$SPH, particle shifting, and tensile instability control, are readily incorporated into the new framework. Finally, the WBSPH formulation is extended to three dimensions and implemented on GPU, with optimization strategies tailored to GPU architectures.

This paper is structured as follows: Section 2 reviews the conventional weakly compressible SPH and presents the construction and analysis of the proposed WBSPH formulation. Section 3 introduces the GPU implementation of WBSPH. Section 4 applies the WBSPH to a series of benchmark cases and the final section provides the concluding remarks.

\section{Construction of SPH formulation with exact hydrostatic balance}
In this section, the conventional weakly compressible SPH formulation is briefly reviewed, and the well-balanced SPH formulation is derived.
\subsection{Conventional weakly compressible SPH}
Weakly compressible SPH is based on the NS equations in Lagrangian form, which are
\begin{equation} \label{eq:nslag}
    \left \{
\begin{aligned}
    \frac{D \rho}{D t} &= -\rho \nabla \cdot \mathbf{U}, \\
    \frac{D \mathbf{U}}{D t} &= -\frac{\nabla p}{\rho} + \nu \nabla^2 \mathbf{U} + \mathbf{G}, \\
    \frac{D \mathbf{x}}{D t} &= \mathbf{U},
\end{aligned}
    \right.
\end{equation}
where, $\rho,\mathbf{U},p, \mathbf{G}, \mathbf{x}$ denote density, velocity, pressure, external force and position, respectively, $\nu = \frac{\mu}{\rho_0}$ is the kinematic viscosity. A weakly compressible equation of state is used to relate pressure with density,
\begin{equation} \label{eq:eos-water}
    p=C_{0}^2(\rho - \rho_{0}),
\end{equation}
where, $\rho_0$ is the reference density. To ensure that the variation of density remains within $1\%$, $C_0$ is chosen to satisfy $C_0 \ge 10|\mathbf{U}|_{max}$, $|\mathbf{U}|_{max}$ denotes the maximum velocity magnitude.

The particle discretizations of spatial differentials in Eq. (\ref{eq:nslag}) are
$$
<\nabla \cdot \mathbf{U}>_i = \sum_j (\mathbf{U}_j - \mathbf{U}_i) \cdot \nabla_i W_{ij} V_j,
$$
$$
<\nabla p>_i = \sum_j (p_i + p_j) \nabla_i W_{ij} V_j,
$$
$$
<\nabla^2 \mathbf{U}>_i = \sum_j (\mathbf{U}_j - \mathbf{U}_i) \frac{(\mathbf{x}_{j} -\mathbf{x}_{i})\cdot \nabla_i W_{ij}}{||\mathbf{x}_{j} -\mathbf{x}_{i}||^2} V_j.
$$
The notation $<>$ denotes the particle approximations in SPH,  $V$ denotes the volume of particles, and $W$ is the smoothing kernel function. C-2 Wendland kernel is used in this work, defined as
\begin{equation} \label{eq:c2kernel}
    W(q) = \left \{
\begin{aligned}
    &\frac{7}{4\pi h^2}(1-\frac{1}{2}q)^4(2q + 1), &q \le 2,\\
    &0, &\text{elsewhere},
\end{aligned}
    \right.
\end{equation}
where, $q = \frac{||\mathbf{x}_j - \mathbf{x}_i||}{h}$, $h$ is the smoothing length, $h = 1.5\Delta x$ in this paper, $\Delta x$ indicates the initial particle space.

\subsection{Well-balanced weakly compressible SPH}
As we discussed before, the well-balanced problem originates from two difficulties: the nonlinearity of the pressure-gradient-over-density term, and the approximation error of gradients evaluated by particle summation. In this section, we will introduce the remedies to these two difficulties, and analyze the well-balanced property of the new formulation.
\subsubsection{Governing equations}
To remove the first difficulty, an auxiliary potential variable $\phi$ is introduced to recast the Lagrangian NS equations, inspired by Chen \cite{wb-gks}, the potential function satisfies the relation
\begin{equation}
    \rho = \rho_0 e^{\frac{\phi}{C_0^2}}.
\end{equation}
It is noteworthy that an auxiliary variable $\alpha$ was defined through $\rho = \rho_0 e^{\frac{\phi}{\alpha}}$ to recast the flux and source terms in NS equations with prescribed potential function $\phi$ in Chen's work \cite{wb-gks}. This is not the case in the present work, however, since the potential function here is not prescribed a priori but is instead treated as an evolving variable determined a posteriori. By this auxiliary potential variable, the NS equations in Eq. (\ref{eq:nslag}) are recast as
\begin{equation} \label{eq:newEqns}
    \left \{
\begin{aligned}
    \frac{D \phi}{D t} &= -C_0^2 \nabla \cdot \mathbf{U}, \\
    \frac{D \mathbf{U}}{D t} &= -\nabla \phi + \nu \nabla^2 \mathbf{U} + \mathbf{G}, \\
    \frac{D \mathbf{x}}{D t} &= \mathbf{U},
\end{aligned}
    \right.
\end{equation}
where the nonlinear pressure-gradient-over-density term $\frac{\nabla p}{\rho}$ is rewritten as the gradient of the potential function $\nabla \phi$, and $\rho \nabla \cdot \mathbf{U}$ is rewritten as $C_0^2 \nabla \cdot \mathbf{U}$.

\subsubsection{A stable and first-order consistent particle approximation of gradient operator}\label{sec:parAppr}
The differential and conservative formulas are widely used to discretize the gradient of a scalar variable in conventional SPH formulation, \\
differential formula:
$$
<\nabla \phi>_i = \sum_j (\phi_j - \phi_i) \nabla_i W_{ij} V_j,
$$
conservative formula:
$$
<\nabla \phi>_i = \sum_j (\phi_j + \phi_i) \nabla_i W_{ij} V_j.
$$
It is well known that the conservative formula can not restore even zeroth-order consistency when the particles are disordered, and the differential formula suffers from the stability issue. Godunov-SPH was proposed \cite{gsph1} and developed \cite{gsph2} to improve the stability of SPH with Riemann solvers. 

In this study, to remove the second difficulty, a stable and first-order consistent particle approximation for the gradient of a scalar variable is proposed, specifically, a Riemann-based particle approximation with kernel correction is constructed to discretize the potential gradient $\nabla \phi$ with first-order consistency, which is formulated as
\begin{equation}
    <\nabla \phi>_i = \sum_j 2(\phi_{ij}^* - \phi_i) \mathbb{L}^{-1} \nabla_i W_{ij} V_j.
    \label{eq:parGrad}
\end{equation}
The renormalization matrix $\mathbb{L}$ is defined by
\begin{equation} \label{eq:renorm}
    \mathbb{L} = \sum_j (\mathbf{x}_j - \mathbf{x}_i) \otimes \nabla_i W_{ij} V_j.
\end{equation}
and $\phi_{ij}^*$ is calculated by Lax-Friedrichs formula
\begin{equation} \label{eq:lfsover}
    \phi_{ij}^* = \frac{\phi_i + \phi_j}{2} - \frac{C_0}{2}(U_{j,n} - U_{i,n}),
\end{equation}
$$
U_n = \frac{\mathbf{U}\cdot (\mathbf{x}_{j} -\mathbf{x}_{i})}{||\mathbf{x}_{j} -\mathbf{x}_{i}||}.
$$

We then analyze the first-order consistency and dissipation property of the particle approximation Eq. (\ref{eq:parGrad}). Substitute Eq. (\ref{eq:lfsover}) into Eq. (\ref{eq:parGrad}),
\begin{equation} \label{eq:numPhiGrad}
    <\nabla \phi>_i = \sum_j (\phi_j - \phi_i)\mathbb{L}^{-1} \nabla_i W_{ij} V_j - C_0 \sum_j (U_{j,n} - U_{i,n}) \mathbb{L}^{-1} \nabla_i W_{ij} V_j.
\end{equation}
Denote the first right-hand-term of Eq. (\ref{eq:numPhiGrad}) as $\widetilde{\nabla \phi}_i$, and consider the Taylor expansion of the potential variable
$$
\phi_j = \phi_i + \nabla \phi_i \cdot (\mathbf{x}_{j} -\mathbf{x}_{i}) + O(h^2)
$$
and substitute it into the first right-hand-term of Eq. (\ref{eq:numPhiGrad})
\begin{equation}
    \begin{aligned}
    \widetilde{\nabla \phi}_i &= \sum_j (\nabla \phi_i \cdot (\mathbf{x}_{j} -\mathbf{x}_{i}) + O(h^2))\mathbb{L}^{-1} \nabla_i W_{ij} V_j \\
    &= \nabla \phi_i \cdot \sum_j (\mathbf{x}_{j} -\mathbf{x}_{i}) \mathbb{L}^{-1} \nabla_i W_{ij} V_j + O(h^2) \\
    &=\nabla \phi_i + O(h^2).
\end{aligned}
\nonumber
\end{equation}
Denote the second right-hand-term of Eq. (\ref{eq:numPhiGrad}) as $\Pi_i$,
\begin{equation}
    \begin{aligned}
        \Pi_i &= C_0 \sum_j (U_{j,n} - U_{i,n}) \mathbb{L}^{-1} \nabla_i W_{ij} V_j \\
        &=C_0 \sum_j (\mathbf{U}_j - \mathbf{U}_i) \cdot \frac{\mathbf{x}_j - \mathbf{x}_i}{||\mathbf{x}_j - \mathbf{x}_i||} \mathbb{L}^{-1} \nabla_i W_{ij} V_j.
    \end{aligned}
    \nonumber
\end{equation}
which is similar to the artificial viscosity proposed in \cite{monaghan2005}, whose formula is
$$
\Pi_i^{art} = \alpha hC_0 \sum_j (\mathbf{U}_j - \mathbf{U}_i) \cdot \frac{\mathbf{x}_j - \mathbf{x}_i}{||\mathbf{x}_j - \mathbf{x}_i||^2} \nabla_i W_{ij} V_j.
$$

Above all, the particle approximation Eq. (\ref{eq:parGrad}) restore first-order consistency when ignoring the artificial viscosity induced by the Riemann solver.

\subsection{Stability improvement techniques} \label{sec:stableImp}
The widely used stability improvement techniques, e.g. $\delta-$SPH, particle shifting techniques and tensile instability control, can be readily extended to the new framework based on Eq. (\ref{eq:newEqns}).

To further improve the stability and alleviate the potential fluctuation, numerical diffusive term is added in the potential equation as that in $\delta-$SPH \cite{deltasph},
$$
D^{\phi}_i = \delta h C_0 \sum_j \Phi_{ij} \cdot \nabla_i W_{ij} V_j,
$$
where,
$$
\Phi_{ij} = 2(\phi_j - \phi_i)\frac{\mathbf{x}_j - \mathbf{x}_i}{||\mathbf{x}_j - \mathbf{x}_i||^2},
$$
and $\delta = 0.1$ in this work.

To regularize the particle distribution, the Particle Shifting Technique (PST) applied in \cite{deltaplusesph} is employed. The velocity deviation arising from irregular particle configurations is expressed as:
\begin{equation}
    \delta \mathbf{U}_i = \text{min}\{ ||\delta \mathbf{U}^*_i||, \frac{U_{max}}{2}\} \frac{\delta \mathbf{U}^*_i}{||\delta \mathbf{U}^*_i||}.
\end{equation}
where, $U_{max}$ denotes the maximum velocity within the computational domain, and the term $\delta \mathbf{U}^*_i$ is evaluated as follows:
$$
\delta \mathbf{U}^*_i = -2h_iU_{max}\sum_j[1+R(\frac{W_{ij}}{W(\Delta x)})^n]\nabla_iW_{ij} V_j.
$$
The constants $R$ and $n$ are set to 0.2 and 4, respectively, as recommended by \cite{deltaplusesph}.

The renormalization matrix Eq. (\ref{eq:renorm}) is also used to detect the free surface particles by its minimum eigenvalue $\lambda_{min}$ without any further effort. The particles with $\lambda_{min} \leq 0.75$ are regarded as free surface particles, which include the free surface and non-uniformly spread particles as in the reference \cite{surfDetect}. The renormalization matrix $\mathbb{L}$ defined in Eq. (\ref{eq:renorm}) may be singular and singular for flows with violent free surface deformation. In this context, a more robust strategy is to combine the formula Eq. (\ref{eq:parGrad}) with the widely-used conservative form, 
\begin{equation}
    <\nabla \phi>_i = 
    \left \{
\begin{aligned}
    &\sum_j (\phi_j + \phi_i) \nabla_i W_{ij} V_j, \qquad &\lambda_{min} < 0.75\\
    &\sum_j 2(\phi_{ij}^* - \phi_i) \mathbb{L}^{-1} \nabla_i W_{ij} V_j, \qquad &otherwise.
\end{aligned}
    \right.
\end{equation}
which is consistent with the idea of tensile instability control (TIC) technique proposed in \cite{tic}. But in hydrostatic equilibrium situations, Eq. (\ref{eq:parGrad}) is used to discretize the gradient of potential variable to restore the well-balanced property.

Above all, the final formulations of the SPH are
\begin{equation} \label{eq:sph}
    \left \{
\begin{aligned}
    \frac{D \phi_i}{D t} &= -C_0^2\sum_j (\mathbf{U}_j - \mathbf{U}_i) \cdot \nabla_i W^c_{ij} V_j + D^{\phi}_i, \\
    \frac{D \mathbf{U}_i}{D t} &= -\sum_j 2(\phi_{ij}^* - \phi_i) \nabla_i W^c_{ij} V_j + \mathbf{G}, \\
    \frac{D \mathbf{x}_i}{D t} &= \mathbf{U}_i + \delta \mathbf{U}_i,
\end{aligned}
    \right.
\end{equation}
where $\nabla_i W^c_{ij}$ is the corrected kernel gradients, defined by $\nabla_i W^c_{ij} = \mathbb{L}^{-1} \nabla_i W_{ij}$, where renormalization matrix $\mathbb{L}$ is defined in Eq. (\ref{eq:renorm}).

For time integration, the strong stability preserving Runge-Kutta second-order (SSP-RK2) method is used in this study, which is\\
First stage:
\begin{equation}
    \left \{
\begin{aligned}
    \phi^{(1)}_i &= \phi^n_i + \Delta t \frac{D\phi_i^{n}}{Dt}, \\
    \mathbf{U}^{(1)}_i &= \mathbf{U}^n_i + \Delta t \frac{D\mathbf{U}_i^{n}}{Dt}, \\
    \mathbf{x}^{(1)}_i &= \mathbf{x}^{n}_i + \Delta t (\mathbf{U}^n_i + \delta \mathbf{U}^n_i). 
\end{aligned}
    \right.
\end{equation}
Second stage:
\begin{equation}
    \left \{
\begin{aligned}
    \phi^{n+1}_i &= \frac{1}{2}\phi^n_i + \frac{1}{2}(\phi^{(1)}_i + \Delta t \frac{D\phi_i^{(1)}}{Dt}), \\
    \mathbf{U}^{n+1}_i &= \frac{1}{2}\mathbf{U}^n_i + \frac{1}{2}(\mathbf{U}^{(1)}_i + \Delta t \frac{D\mathbf{U}_i^{(1)}}{Dt}), \\
    \mathbf{x}^{n+1}_i &= \frac{1}{2}\mathbf{x}^n_i + \frac{1}{2}(\mathbf{x}^{(1)}_i + \Delta t (\mathbf{U}^{(1)}_i + \delta \mathbf{U}^{(1)}_i)). \\
\end{aligned}
    \right.
\end{equation}
Time step, $\Delta t$, is obtained as the minimum over the following three conditions:
\begin{equation}
    \left \{
\begin{aligned}
    &\Delta t_c = \frac{\Delta x}{C_0+||\mathbf{U}||}, \Delta t_d = \frac{\Delta x^2}{2 \nu}, \Delta t_g = \sqrt{\frac{2 \Delta x}{||\mathbf{G}||}}, \\
    &\Delta t = \mathrm{CFL}\min (\Delta t_c, \Delta t_d, \Delta t_g),
\end{aligned}
    \right.
\end{equation}
where $\mathrm{CFL}$ is the Courant-Friedrichs-Lewy number, and $\mathrm{CFL} = 0.2$ throughout this work.

\subsection{Wall boundary treatment}
In this study, the solid walls are discretized with fixed dummy particles to complete the support of the smoothing kernel near the wall boundary, as introduced in \cite{bnCond}. The attributes of each particle are determined from the nearest fluid particle, e.g. for non-slip wall boundaries,
\begin{equation} \label{eq:phiBn}
    \phi_d = \phi_f + \mathbf{G} \cdot (\mathbf{x}_d - \mathbf{x}_f),
\end{equation}
\begin{equation} \label{eq:velBn}
    \mathbf{U}_d = 2\mathbf{U}_w - \mathbf{U}_f,
\end{equation}
where subscript $d$ indicates dummy particles, $f$ indicates neighboring fluid particles and $w$ indicates the wall boundary. For hydrostatic equilibrium state, the treatment for potential variable Eq. (\ref{eq:phiBn}) can restore the analytical solution exactly, which will be discussed in the next section. 

The particle space is always much larger than the boundary layer thickness for most cases, in this situation, the non-slip velocity condition Eq. (\ref{eq:velBn}) and the viscous interaction of a fluid particle with adjoint dummy particle will enlarge the viscous effect severely. In this study, the slip wall boundary is imposed and viscous interactions with dummy particles are neglected as most SPH simulations \cite{bnCond,bapr}. The condition for potential variable Eq. (\ref{eq:phiBn}) still holds for slip wall, but the velocity condition becomes
\begin{equation}
    \mathbf{U}_d = \mathbf{U}_f - (\mathbf{U}_f \cdot \mathbf{n}_d)\mathbf{n}_d,
\end{equation}
where $\mathbf{n}_d$ is the unit normal vector at the dummy particle.

\subsection{Well-balanced property}
In this work, we focus on the hydrostatic equilibrium situation with any form of bottom topographies under constant gravitational force, which satisfies
\begin{equation} \label{eq:hydrostatic}
    \frac{\nabla p}{\rho} = \mathbf{G}.
\end{equation}
Combined with the linearized equation of state Eq. (\ref{eq:eos-water}), the analytical solution is
$$
\rho = \rho_0 e^{\frac{\mathbf{G}\cdot \mathbf{x}-H ||\mathbf{G}||}{C_0^2}},
$$
where $H$ is the altitude of the quiescent water level. The hydrostatic solutions for the potential variable and velocity satisfy
\begin{equation} \label{eq:eqSol}
    \left \{
\begin{aligned}
    \phi &= \mathbf{G}\cdot \mathbf{x}-H ||\mathbf{G}||, \\
    \mathbf{U} &= \mathbf{0}.
\end{aligned}
    \right.
\end{equation}
For the hydrostatic equilibrium situation under constant gravitational force, $H$ and $\mathbf{G}$ are constants, then the potential $\phi$ is a linear function for hydrostatic solution. The particle approximation of $\nabla \phi$ constructed in Section \ref{sec:parAppr} reproduce the gravitational force exactly for hydrostatic state, so that the hydrostatic solution Eq. (\ref{eq:eqSol}) satisfies the discretization of the present SPH formulations Eq. (\ref{eq:sph}).

\section{GPU implementation}
There are three main steps in SPH workflow: neighboring particles searching, force calculation and time integration. To fully exploit the computational power of GPU architectures, all these three steps are fully executed on GPU. Like most of the GPU implementations, one particle is manipulated by one CUDA thread \cite{gpu-splitMegre,gpu-early1}. The realization of these three main parts and some other important steps are illustrated in this section.

The particle attributes, such as velocity, position, forces, etc, are arranged into "Structure of Array" (SoA), which ensures memory coalescence for maximal performance. Single precision is adopted in this study out of the consideration of hardware performance and memory bound of GPU platform, but we still use the double precision for well-balanced testing cases. To mitigate the effect of the accumulation of round-off error, we use the Kahan summation \cite{kahan} in the time integration step to update the particle attributes.

\begin{figure}[!htbp]
    \centering
    \includegraphics[width=0.8\linewidth]{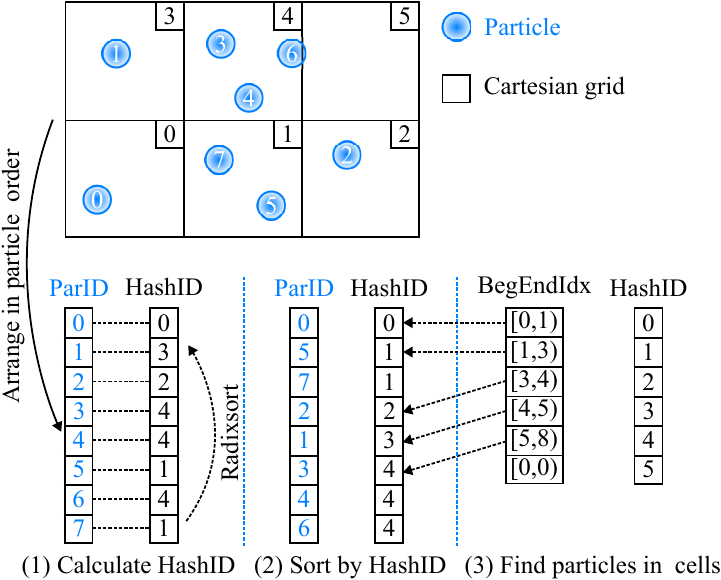}
    \caption{Schematic of neighboring list searching algorithm: (1) Calculate the HashID of each particle (2) Sort the particles using radixsort algorithm by HashID array (3) Find the begin and end particle index inside each Cartesian cell.}
    \label{fig:fig1}
\end{figure}

In the neighbor-list searching stage, we divide the computational domain by background Cartesian grid at the size of $R = 2 \times h_{max}$, which is the maximum smoothing length of all the particles, the neighboring particles only appear in the neighboring Cartesian cells, which reduces the computational complexity from $O(N^2)$ to $O(N)$. The complete searching process consists of three steps, as shown in Figure \ref{fig:fig1}: 
\begin{enumerate}
    \item Calculate the HashID of each particle, which is defined as
    \begin{equation}
        HashID = i_x + i_y \times N_x + i_z \times N_x \times N_y,
        \label{eq:hashid}
    \end{equation}
    \begin{equation}
        i_x = \lfloor \frac{x - x_{min}}{R} \rfloor,i_y = \lfloor \frac{y - y_{min}}{R} \rfloor,i_z = \lfloor \frac{z - z_{min}}{R} \rfloor,
        \label{eq:ixyz}
    \end{equation}
    $$
    N_x = \lfloor \frac{x_{max} - x_{min}}{R} \rfloor,N_y = \lfloor \frac{y_{max} - y_{min}}{R} \rfloor,N_z = \lfloor \frac{z_{max} - z_{min}}{R} \rfloor.
    $$
    where $(x_{min},y_{min},z_{min})$ and $(x_{max},y_{max},z_{max})$ are the minimal and maximal extents of the computational domain, $\lfloor \rfloor$ is the flooring function.
    \item Sort the HashID and ParID array in ascending sequence of HashID array by radixsort algorithm \cite{radixsort}. After the sorting operation, the particles with the same HashID will be stored continuously in memory, which benefits memory coalescence.
    \item Collect the begin and end particle index for each Cartesian cells. 
\end{enumerate}
With the above three steps, Algorithm \ref{alg:alg1} shows the complete neighboring particles searching process, and similar algorithm is applied in \cite{gpu-early1} and \cite{green}.

\begin{algorithm}[!htbp] 
\caption{Neighbor-list searching algorithm}
\label{alg:alg1}
\begin{algorithmic}[1]
\State Calculate $i_x,i_y,i_z$ by Eq. (\ref{eq:ixyz})
\For{$k = -1$ \textbf{to} $1$}
    \State $i_{z,nb} = i_z + k$
    \For{$j = -1$ \textbf{to} $1$}
        \State $i_{y,nb} = i_y + j$
        \For{$i = -1$ \textbf{to} $1$}
            \State $i_{x,nb} = i_x + i$
            \State Calculate the $HashID_{nb}$ by Eq. (\ref{eq:hashid})
            \State Get the start and end particle index from $BegEndIdx$ array
            \For{$ipar = BegIdx$ \textbf{to} $EndIdx$}
                \State Calculate distance $d$ against particle $ipar$
                \If{$d \le 2*h$}
                    \State Calculate particle interactions
                \EndIf
            \EndFor
        \EndFor
    \EndFor
\EndFor
\end{algorithmic}
\end{algorithm}

In the particle interaction stage, two simplifications are introduced to reduce the computational cost on GPU architectures: a simplified kernel correction and an estimate of the minimum eigenvalue of the renormalization matrix. Their detailed implementations are described below.

Firstly, an equivalent form to the particle approximation Eq. (\ref{eq:parGrad}) is proposed to move the kernel correction outside the summation operator,
$$
\mathbb{L}^{T} <\nabla \phi>_i = \sum_j 2(\phi_{ij}^* - \phi_i) \nabla_i W_{ij} V_j,
$$
by which the kernel correction for each neighboring particle is simplified to solve a $2\times2$ (for 2D situation, $3\times3$ for 3D situation) linear system for the final particle summation.

Secondly, for a radial kernel function—such as the C-2 Wendland kernel of Eq. (\ref{eq:c2kernel}) adopted in this work—the renormalization matrix in Eq. (\ref{eq:renorm}) is symmetric, so that its eigenvalues can be evaluated through closed-form expressions \cite{eignvalue} rather than the commonly used iterative procedures, at a considerably lower computational cost.

\section{Numerical examples}
The following examples demonstrate the performance of the proposed well-balanced SPH and its GPU implementation. A series of hydrostatic cases with different bottom topographies is first considered to verify the well-balanced property. The dam break against an obstacle and the water entry of a horizontal cylinder are then presented to illustrate the capability of the method in capturing high-speed impact phenomena. Finally, a full-scale aircraft ditching case is simulated for validation. In all  cases, the reference density is $\rho_0 = 1000kg/m^3$, the kinematic viscosity coefficient is $\nu = 10^{-4} m^2/s$, the gravitational acceleration is $G = -9.81m/s^2$ (y-direction for 2D cases and z-direction for 3D cases), and the artificial sound speed is $C_0 = 30 m/s$. All solid walls are treated as slip boundaries.

\subsection{Hydrostatic equilibrium cases}
In this section, hydrostatic equilibrium states over rectangular, triangular, and Gaussian bottom topographies are considered to assess the well-balanced property of the proposed SPH formulation. A container of size $(x,y) \in [0,1]\times [0,1]m$ is filled with water to an initial depth of $H = 1m$, and the bottom topographies, of height $0.5m$ and width $0.4m$ are placed at the center of the container floor. The $L_2$ error of the velocity magnitude, defined as
$$
L_2 = \sqrt{\frac{\sum_j^{N_p} ||\mathbf{U}_j||^2 V_j}{\sum_j^{N_p} V_j}},
$$
is collected and compared for all the cases.

\begin{figure}[!htbp]
    \centering
    \includegraphics[trim = 10 20 10 40, clip,width=0.6\linewidth]{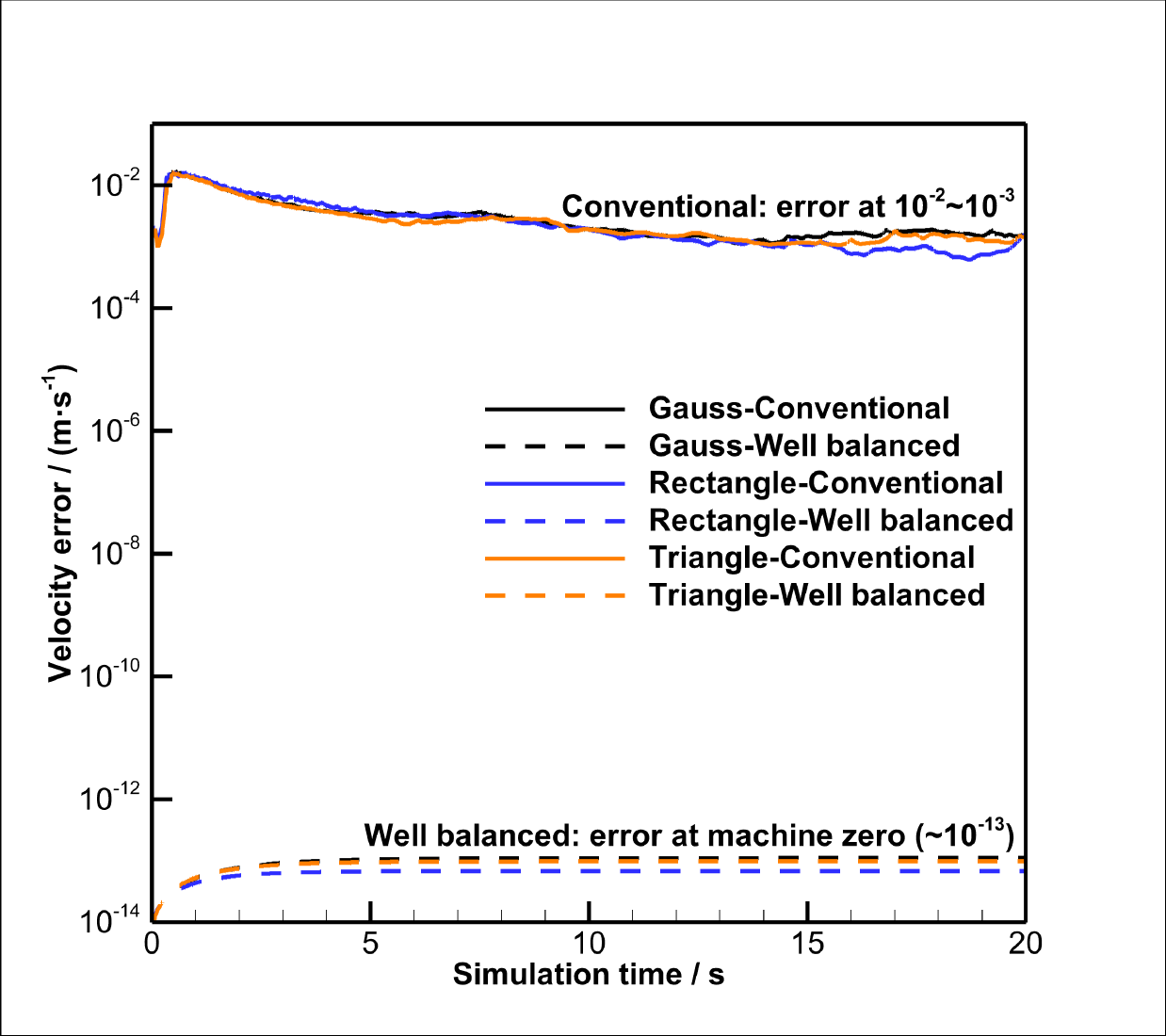}
    \caption{Hydrostatic cases: Comparison of $L_2$ error for hydrostatic equilibrium solutions with rectangular, triangular and Gaussian bottom topographies by WBSPH and conventional weakly compressible SPH. Conventional SPH gives $10^{-2}-10^{-3}$ velocity error while WBSPH achieves machine zero error.}
    \label{fig:fig2}
\end{figure}

The initial particle spacing is $\Delta x = 1 / 100m$, and all the boundaries are treated as slip walls. To restore hydrostatic equilibrium solution, the well-balanced scheme is applied without the stability improvement techniques introduced in Section \ref{sec:stableImp}. Figure \ref{fig:fig2} compares the $L_2$ error of the WBSPH with that of the conventional SPH scheme with artificial viscosity: whereas the conventional scheme yields errors of order $10^{-2}-10^{-3}m/s$, the well-balanced scheme converges to essentially machine-zero error, of order $10^{-13}m/s$.

The left column of Figure \ref{fig:fig3} shows the velocity vectors, colored by contours of the vertical velocity component, at $T = 20s$ for the rectangular, triangular, and Gaussian bottom topographies. The imbalance inherent in conventional SPH clearly induces pronounced spurious flows, most notably in the corners where the free surface intersects the wall boundaries. The right column of Figure \ref{fig:fig3} shows the resulting rise of the free surface driven by these spurious flows, an artifact also reported in \cite{longTermSPH}, which is detrimental to long-term SPH simulations. In contrast, the WBSPH preserves the initial free-surface elevation exactly.

\begin{figure}[!htbp]
	\centering
    \subfigure{\includegraphics[width = 0.9\columnwidth]{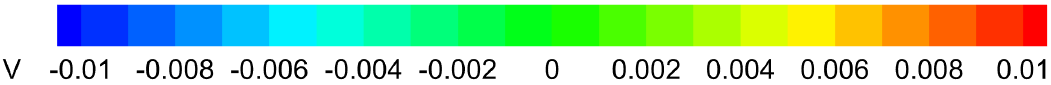}} \\
    \setcounter{subfigure}{0}
	\subfigure[Velocity vector with rectangular bottom]{\includegraphics[trim = 5 5 5 5, clip, width = 0.47\columnwidth]{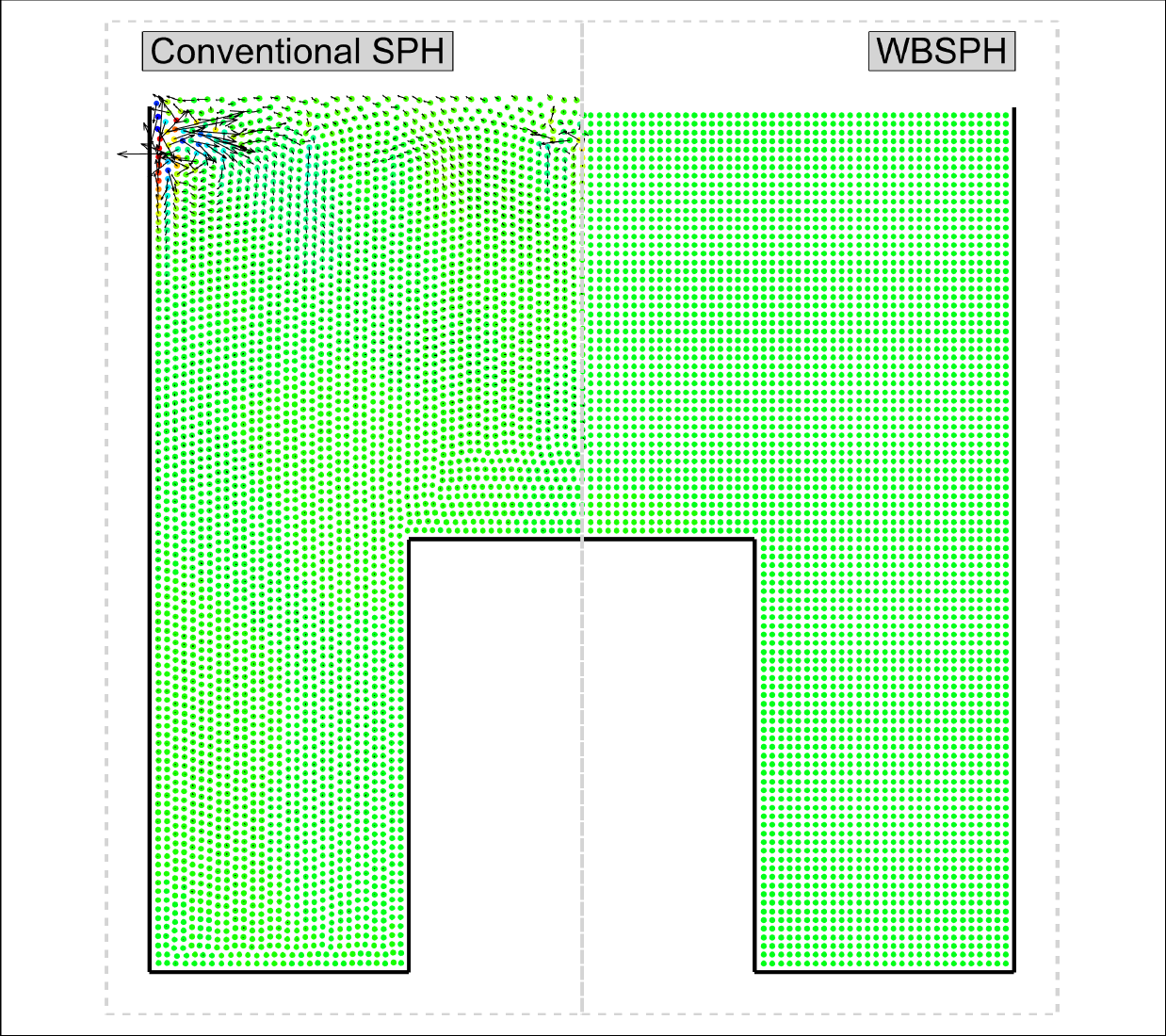}} \hspace{5pt}
	\subfigure[Particle distribution near free surface]{\includegraphics[width = 0.47\columnwidth]{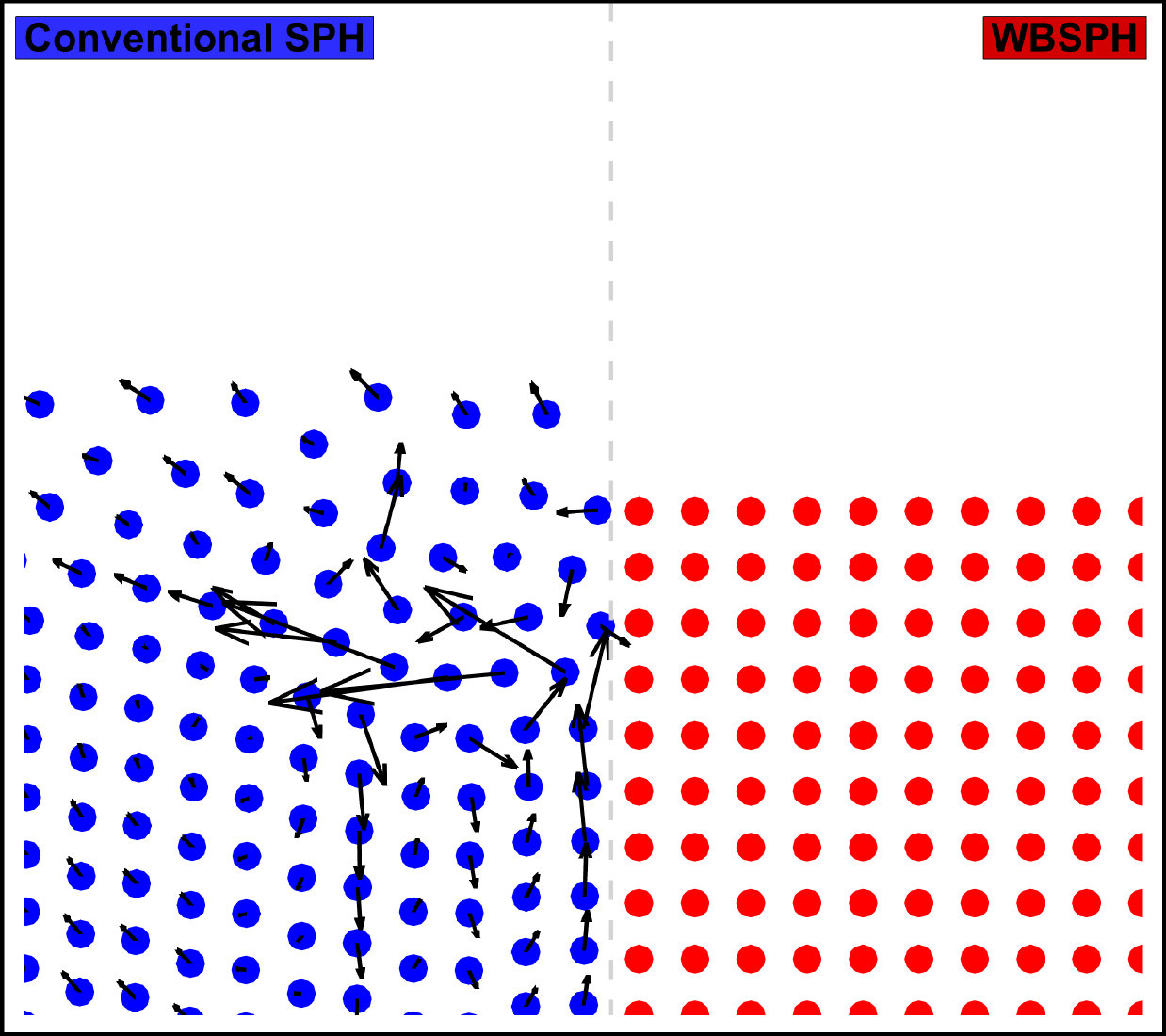}} \\
    \subfigure[Velocity vector with triangular bottom]{\includegraphics[trim = 5 5 5 5, clip, width = 0.47\columnwidth]{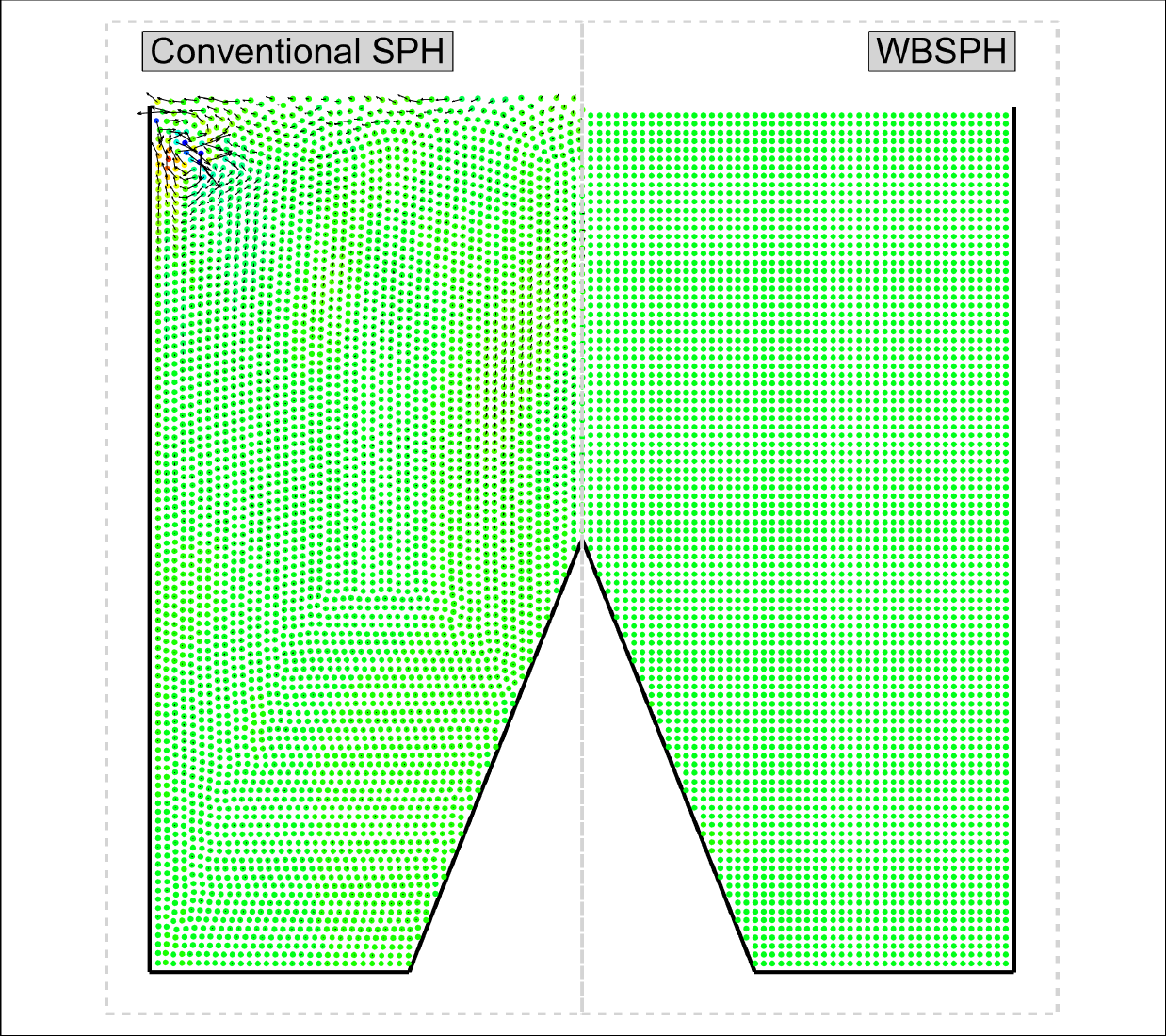}} \hspace{5pt}
	\subfigure[Particle distribution near free surface]{\includegraphics[width = 0.47\columnwidth]{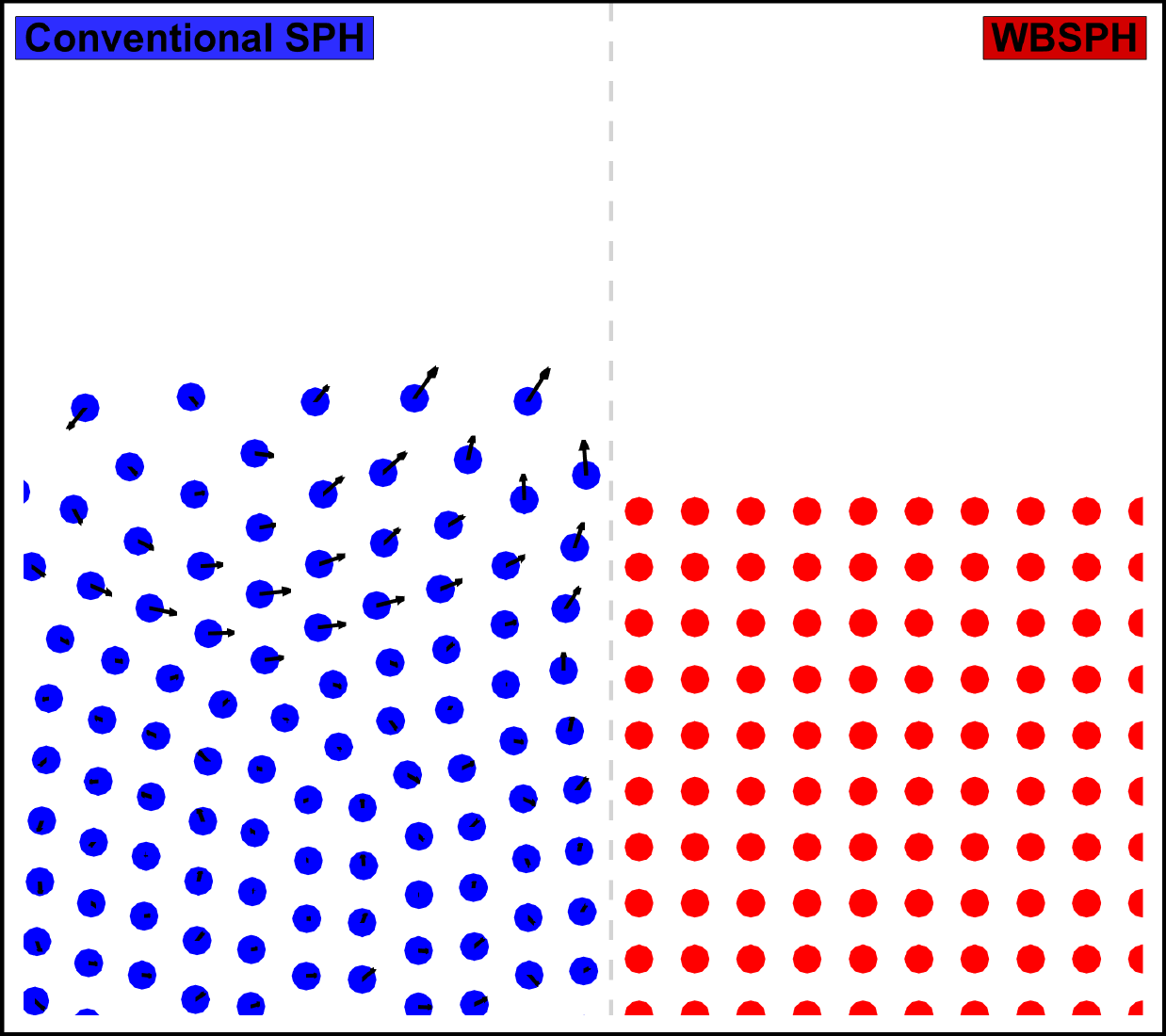}} \\
	\caption{Hydrostatic cases: Comparison of numerical results between WBSPH and conventional SPH. Left column: Velocity vector colored by the contours of V-component velocity, right column: Particle distributions near the free surface at the midline. First row: Rectangular bottom, second row: Triangular bottom, third row: Gaussian bottom.}
	\label{fig:fig3}
\end{figure}

\begin{figure}[!htbp]
    \ContinuedFloat
    \setcounter{subfigure}{4}
	\centering
    \subfigure[Velocity vector with Gaussian bottom]{\includegraphics[trim = 5 5 5 5, clip, width = 0.47\columnwidth]{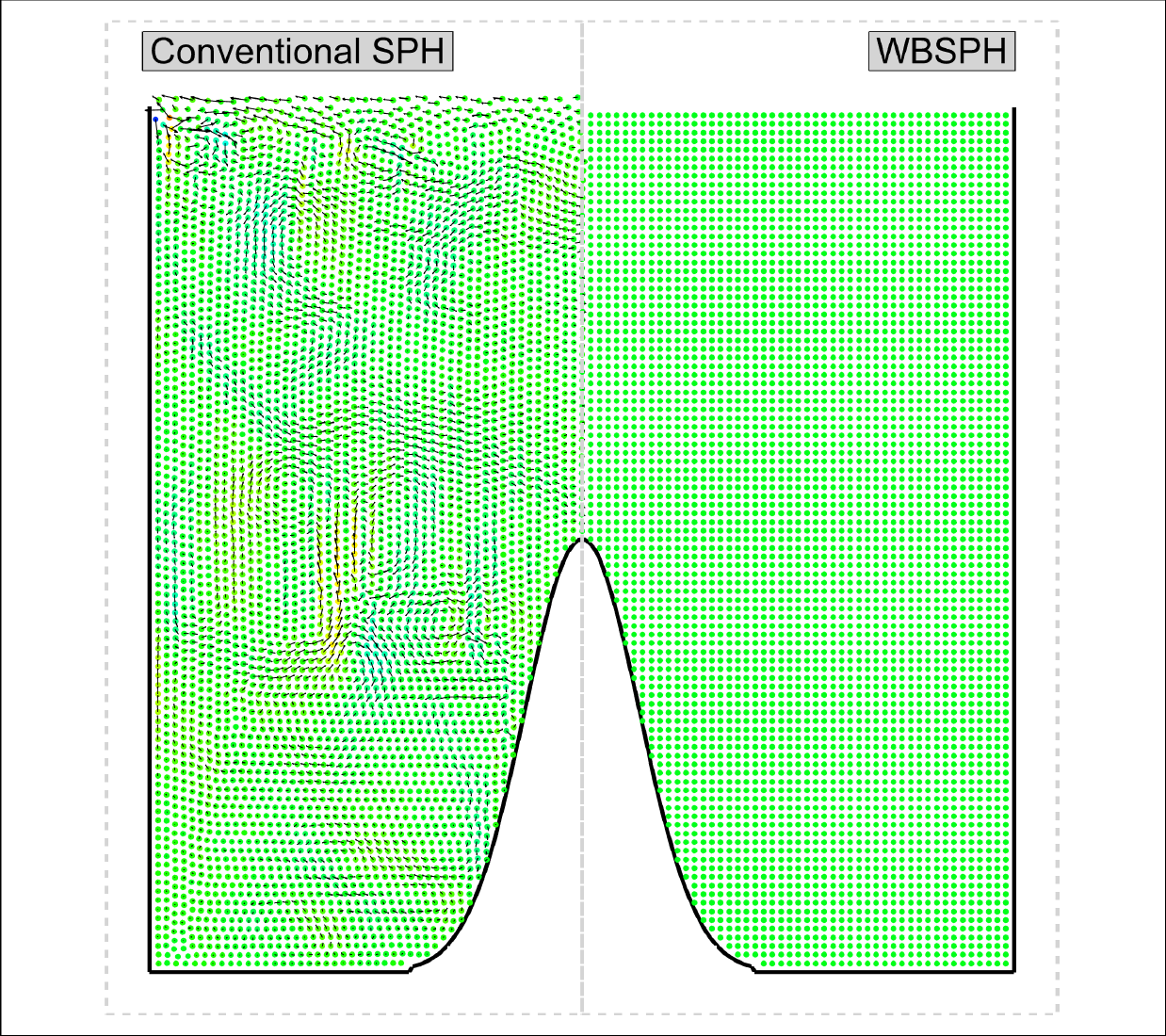}} \hspace{5pt}
	\subfigure[Particle distribution near free surface]{\includegraphics[width = 0.47\columnwidth]{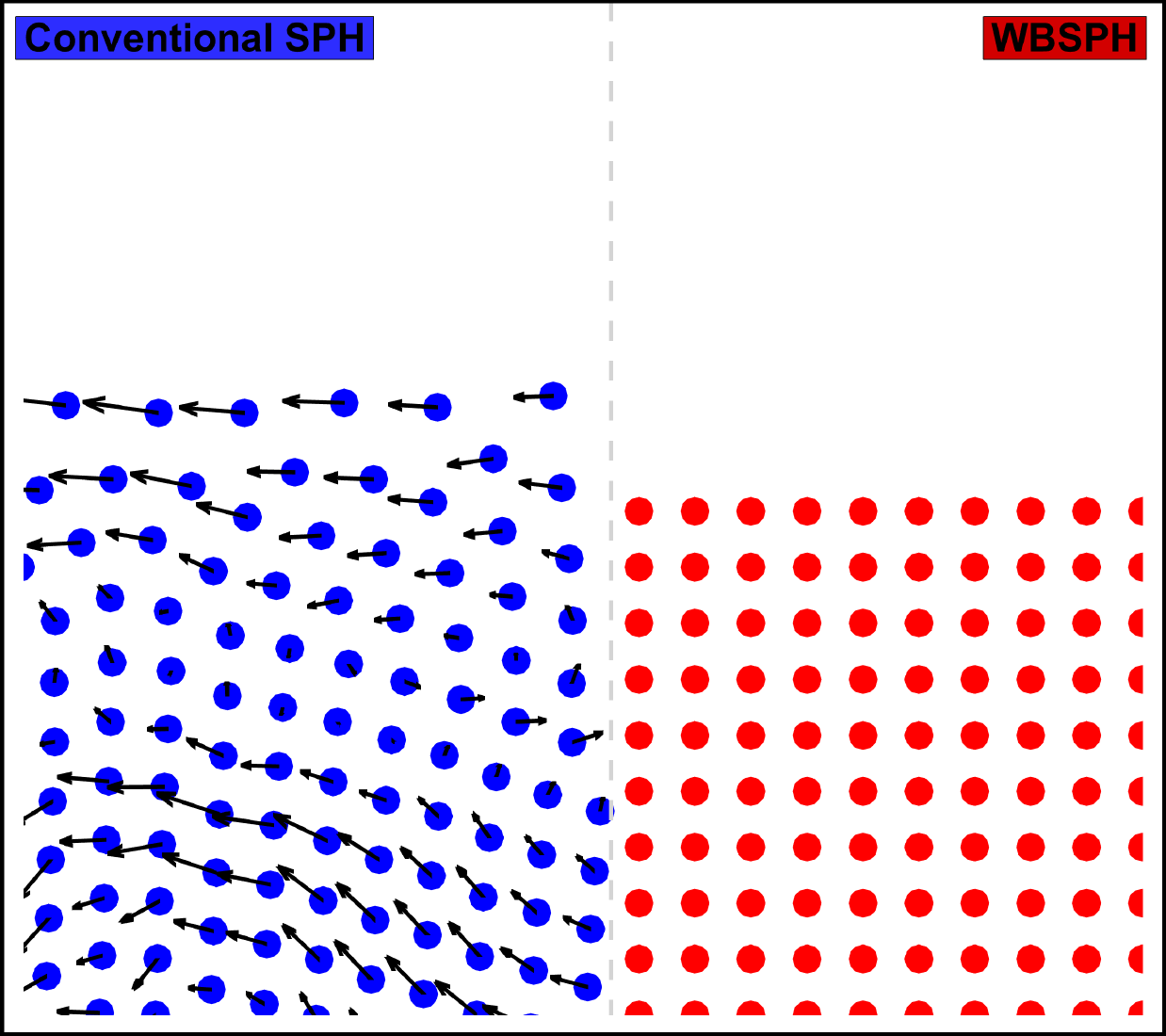}} \\
	\caption{(Continue) Hydrostatic cases: Comparison of numerical results between WBSPH and conventional SPH. Left column: Velocity vector colored by the contours of V-component velocity, right column: Particle distributions near the free surface at the midline. First row: Rectangular bottom, second row: Triangular bottom, third row: Gaussian bottom.}
	\label{fig:fig3-conti}
\end{figure}

\subsection{Dam breaking against an obstacle}
\begin{figure}[!htbp]
	\centering
    \setcounter{subfigure}{0}
	\subfigure[Computational domain]{\includegraphics[width = 0.58\columnwidth]{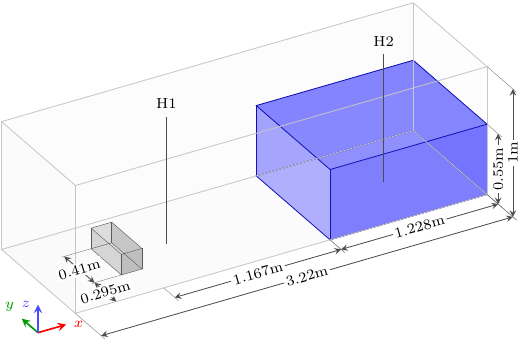}} \hspace{5pt}
	\subfigure[Probe distribution]{\includegraphics[width = 0.38\columnwidth]{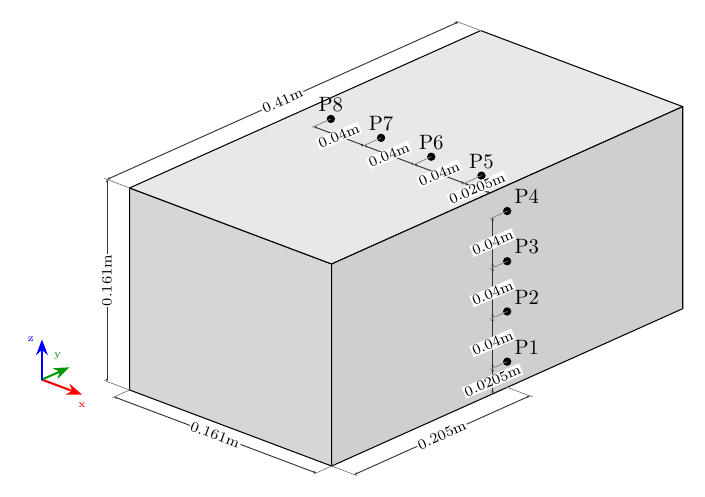}} 
	\caption{Schematic of dam breaking problem with an obstacle. (a) Geometric diagram of the computational domain, (b) Distribution of pressure probes on the surface of the obstacle.}
	\label{fig:fig4}
\end{figure}

The dam breaking problem is a very popular validation case for free surface evolution and high-speed impact, which has detailed experimental data \cite{dambreak-exp}. In this case, a block of water of size $1.228\times 1 \times 0.55m$ is released initially in a reservoir of size $3.22\times 1 \times 2m$. The water flows over the floor of the reservoir and hits the obstacle, as shown in Figure \ref{fig:fig4}.

\begin{figure}[!htbp]
	\centering
	\subfigure[H1]{\includegraphics[trim = 10 20 10 40, clip,width = 0.48\columnwidth]{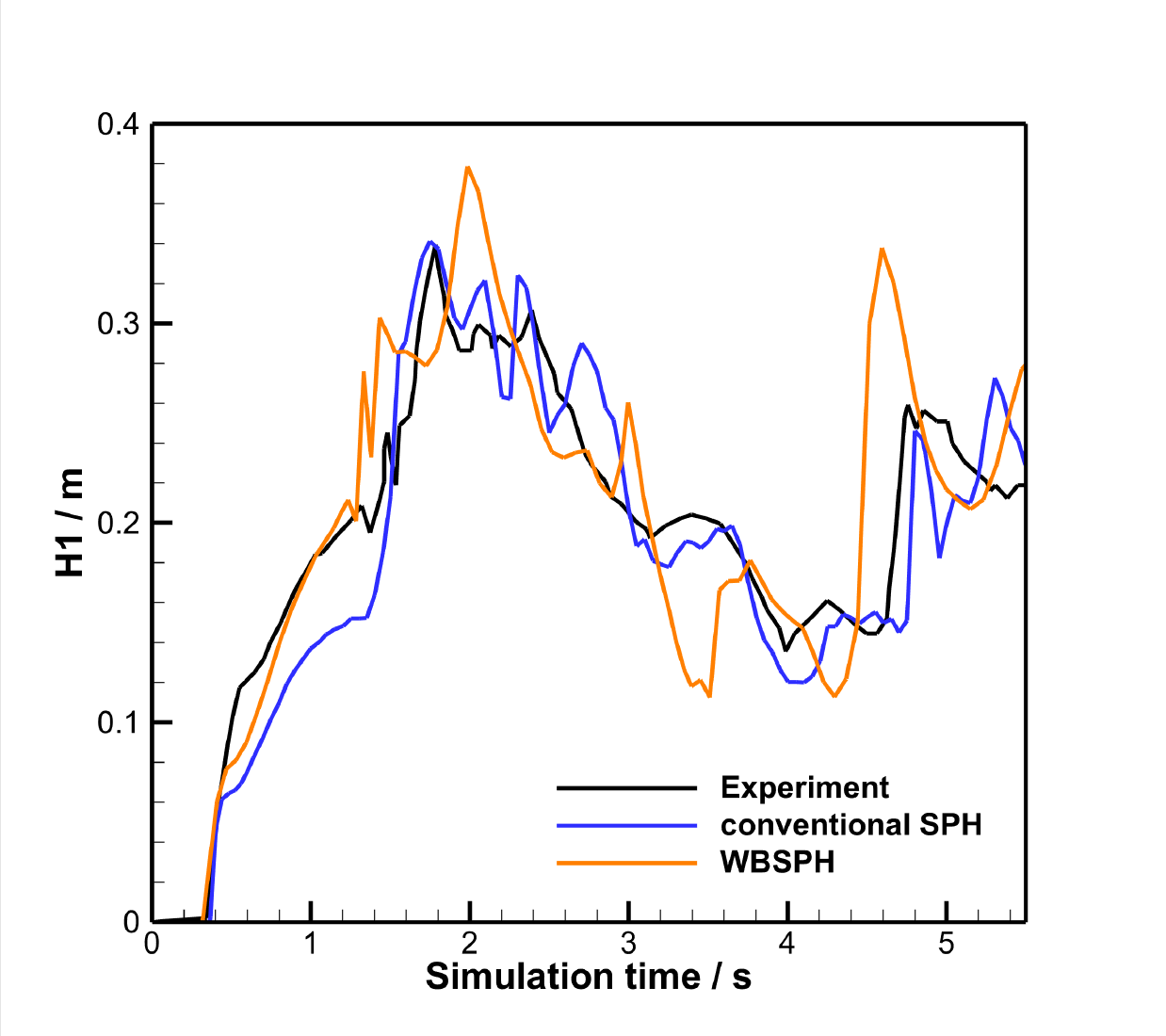}} \hspace{5pt}
	\subfigure[H2]{\includegraphics[trim = 10 20 10 40, clip,width = 0.48\columnwidth]{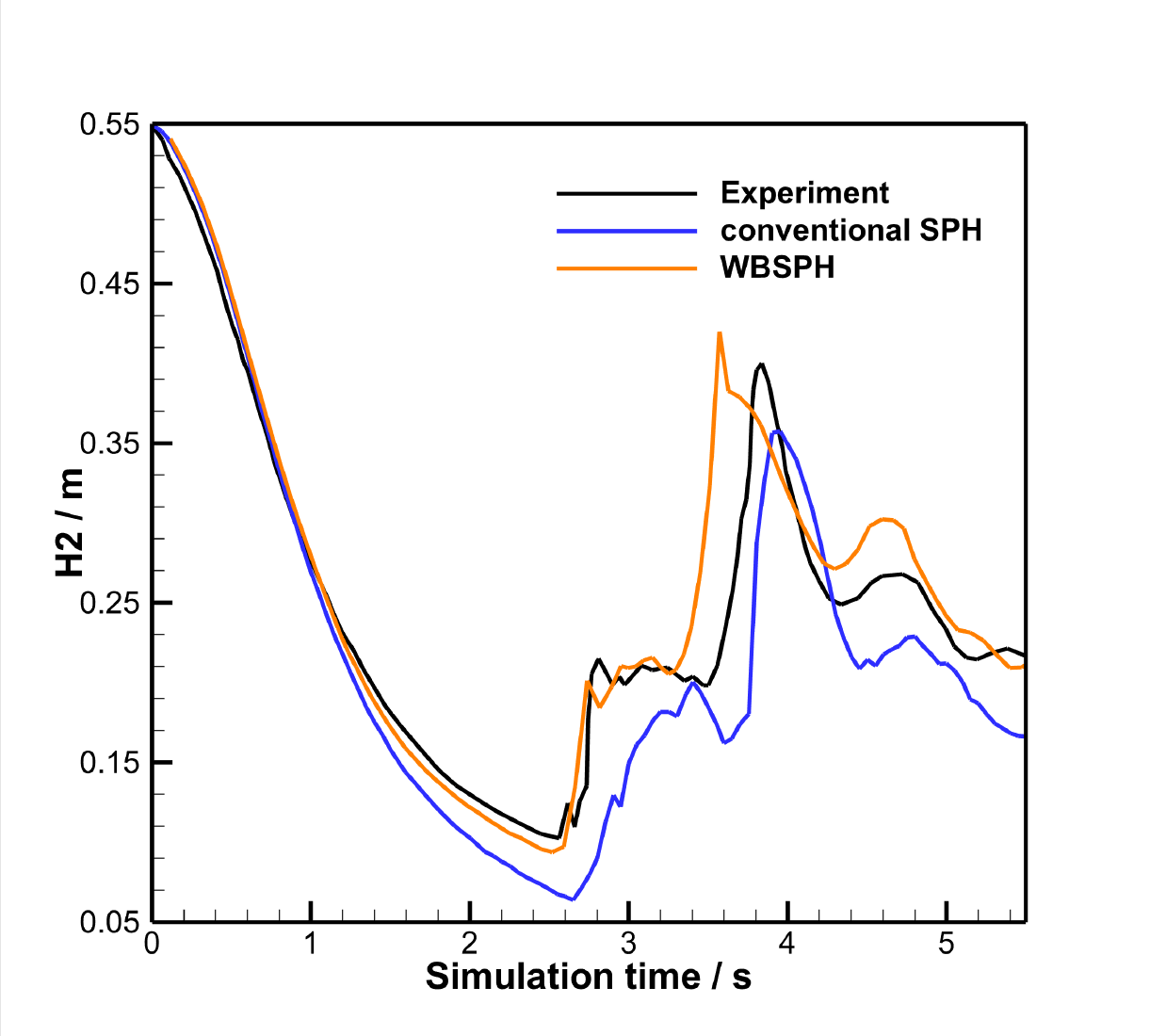}} 
	\caption{Dam breaking problem against an obstacle: Vertical water heights at the probes H1 (in front of the obstacle) and H2 (in the reservoir)}
	\label{fig:fig5}
\end{figure}

The initial particle size is $\Delta x = 0.161 / 30m$, which means the width of the obstacle is discretized by $30$ particles, the total number of particles consists of about 3.29 million boundary particles and 4.30 million fluid particles. Figure \ref{fig:fig5} shows the history of water height at two locations: $H_1$ just in front of the obstacle and $H_2$ in the reservoir, obtained by the present WBSPH and by conventional SPH against the experimental data. In general, good agreements have been achieved by numerical SPH simulations with the experimental data, but conventional SPH systematically underestimates the water depth, whereas the WBSPH result follows the measured curve more closely in both amplitude and phase, particularly before the reflected wave returns at $T=3.4s$.

\begin{figure}[!htbp]
	\centering
	\subfigure[P1]{\includegraphics[trim = 10 20 10 55, clip, width = 0.48\columnwidth]{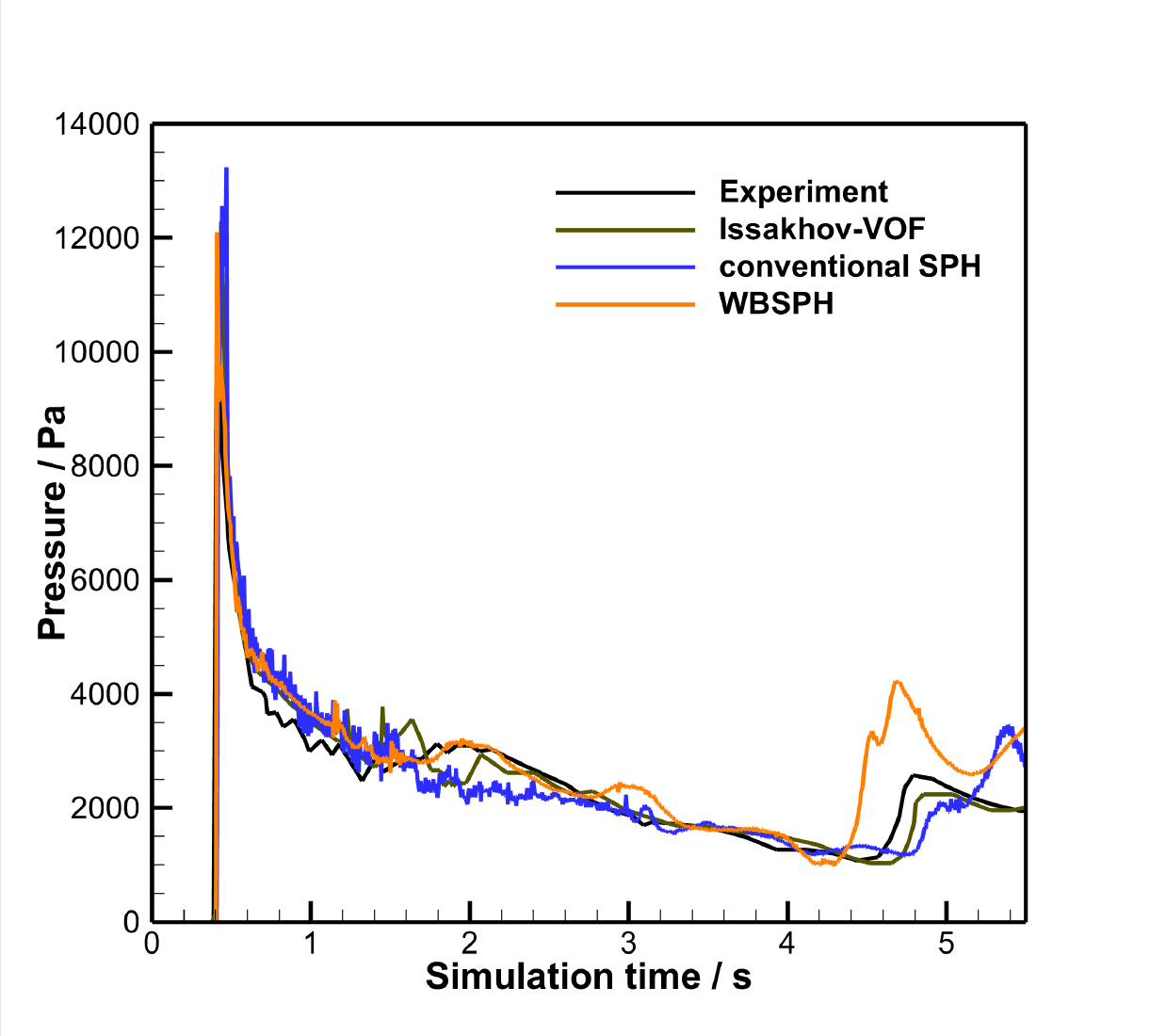}} \hspace{5pt}
	\subfigure[P5]{\includegraphics[trim = 10 20 10 55, clip, width = 0.48\columnwidth]{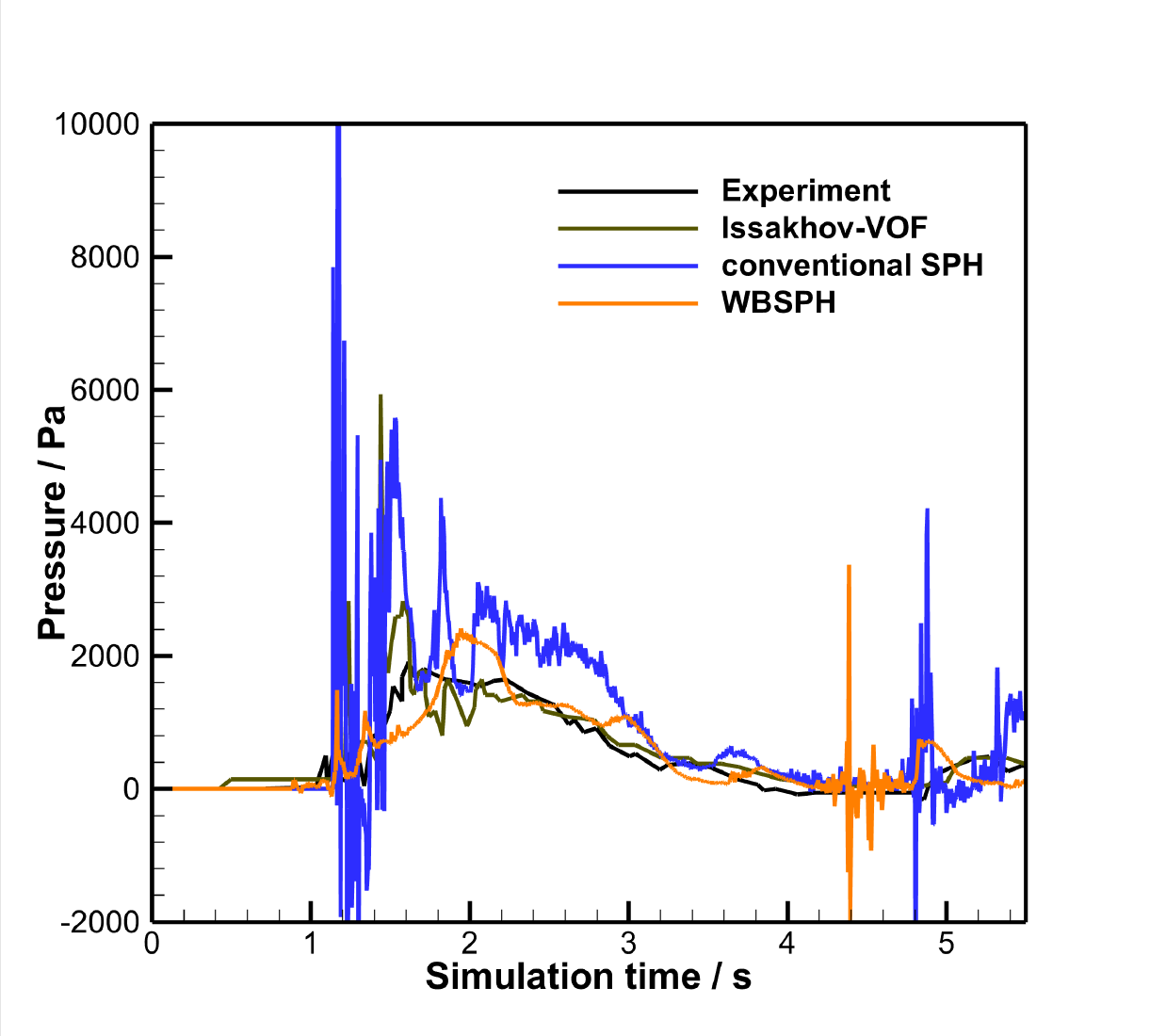}} \\
	\caption{Dam breaking problem against an obstacle: Histories of impact pressure at locations $P_1$ (the lowest on the obstacle) and $P_5$ (at the top of the obstacle). WBSPH gives more consistent pressure against experiment data with much less numerical oscillations than conventional SPH.}
	\label{fig:fig6}
\end{figure}

Figure \ref{fig:fig6} demonstrates the history of pressure at locations $P_1$ and $P_5$, obtained by the WBSPH and by conventional SPH, against the experimental measurements and the VOF results of Issakhov \cite{issakhov}. The WBSPH signal is noticeably smoother and cleaner during the subsequent decay after the first impulsive peak at about $T=0.4s$, following the experimental and VOF curves more closely, whereas the conventional SPH exhibits persistent high-frequency oscillations about the measured trend.

\begin{figure}[!htbp]
	\centering
    \setcounter{subfigure}{0}
	\subfigure[T=0.412s]{\includegraphics[trim = 5 20 15 30, clip, width = 0.48\columnwidth]{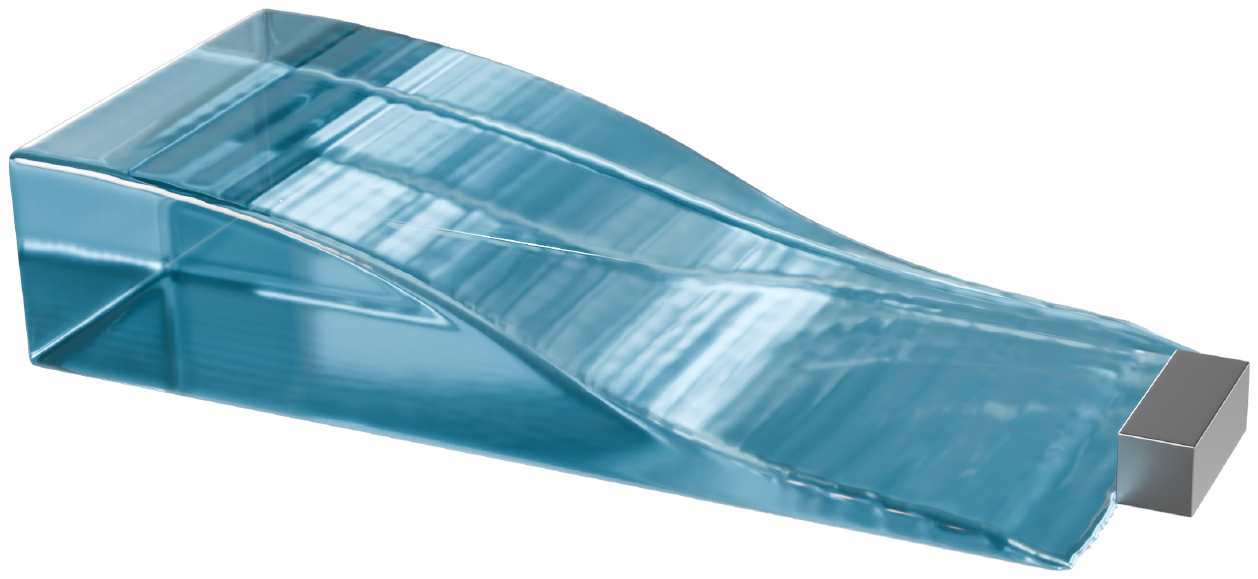}} \hspace{5pt}
	\subfigure[T=1s]{\includegraphics[trim =  5 20 15 30, clip, width = 0.48\columnwidth]{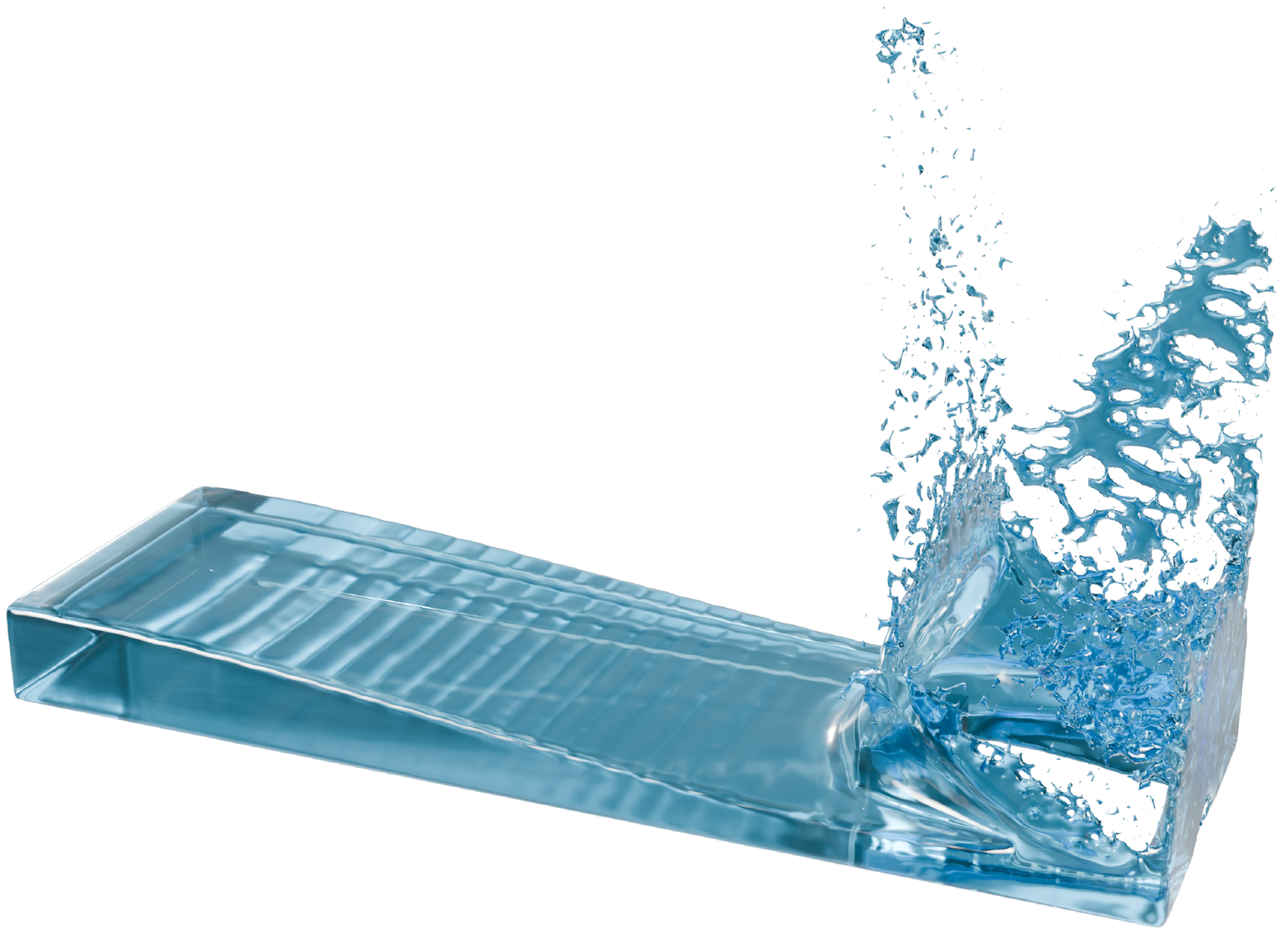}} \\
    \subfigure[T=2s]{\includegraphics[trim =  5 20 15 50, clip, width = 0.48\columnwidth]{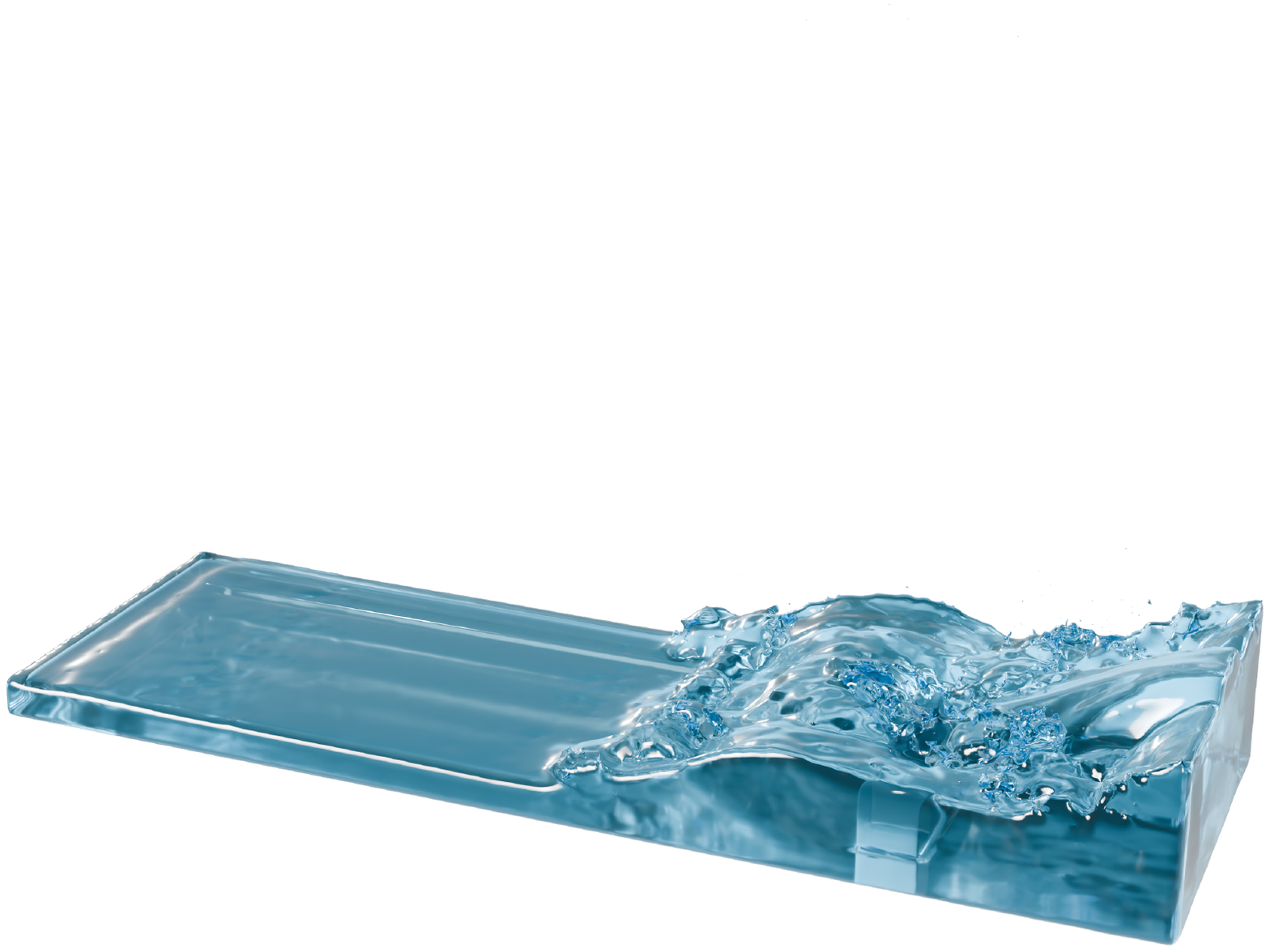}} \hspace{5pt}
	\subfigure[T=4s]{\includegraphics[trim =  5 20 15 50, clip, width = 0.48\columnwidth]{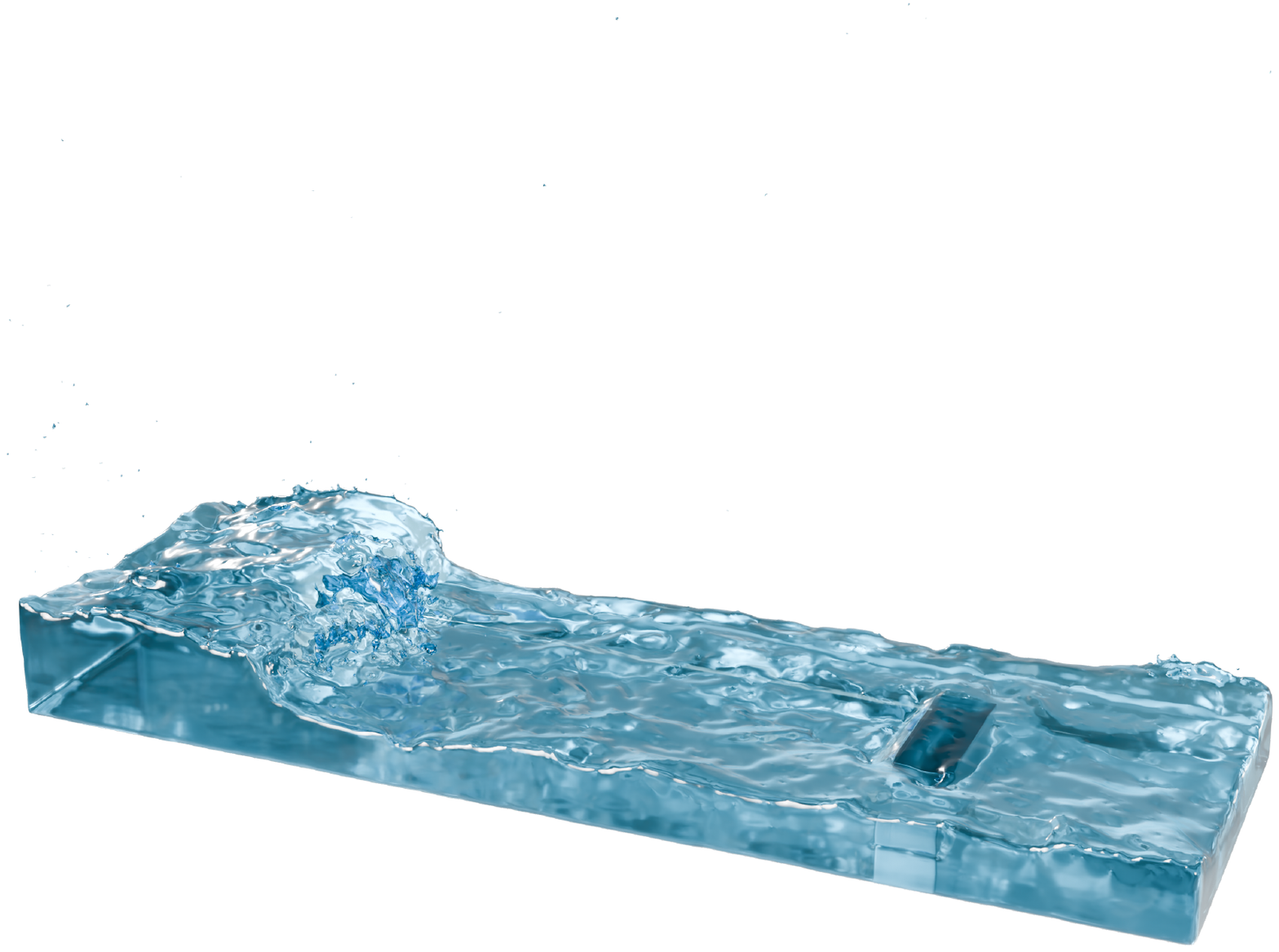}} \\
	\caption{Dam breaking problem against an obstacle: Time evolutions of water body at $T = 0.412, 1, 2, 4s.$}
	\label{fig:fig7}
\end{figure}

Figure \ref{fig:fig7} presents the time evolution of the dam breaking against an obstacle at $T = 0.412$, $1$, $2$, and $4s$. Before $T = 0.412s$, the collapsing water column propagates along the dry bed and develops clean free surface free of spurious particle disorder. At $T = 1s$, the water front hits the obstacle and is deflected into a violent upward jet, accompanied by the ejection of fragmented sheets and scattered droplets; the strongly high-speed impact is captured robustly. At $T = 2s$, the jet has broken up and fallen back onto the underlying flow and the reflected wave begins to travel back towards the upstream wall. Then the reflected wave sloshes back and forth in the reservoir, but progressively decaying to settle down. The proposed WBSPH formulation demonstrates the capability to handle violent impact and complex free-surface fragmentation.

\subsection{Water entry of a horizontal cylinder}
The water entry problem is a classic problem related to many applications and is widely used to validate the robustness and accuracy of numerical methods for high-speed impact problem. In this case, a horizontal cylinder with diameter $D = 0.05m$ and length $L = 0.2m$ falls into a tank with the length of $1m$, width of $0.4m$ and water height of $0.6m$ as the set-up in \cite{sun1}. The initial impact velocity of the cylinder is $V_0 = -6.22m/s$, and the cylinder falls under the balance of gravity force and interaction force with water, and the density of the solid is $\rho_{solid} = 1370kg / m^3$. The initial particle size is $\Delta x = 0.05 / 20m$, the total particle number is about 16 million.

\begin{figure}[!htbp]
    \centering
    \includegraphics[trim = 10 20 10 40, clip,width=0.6\linewidth]{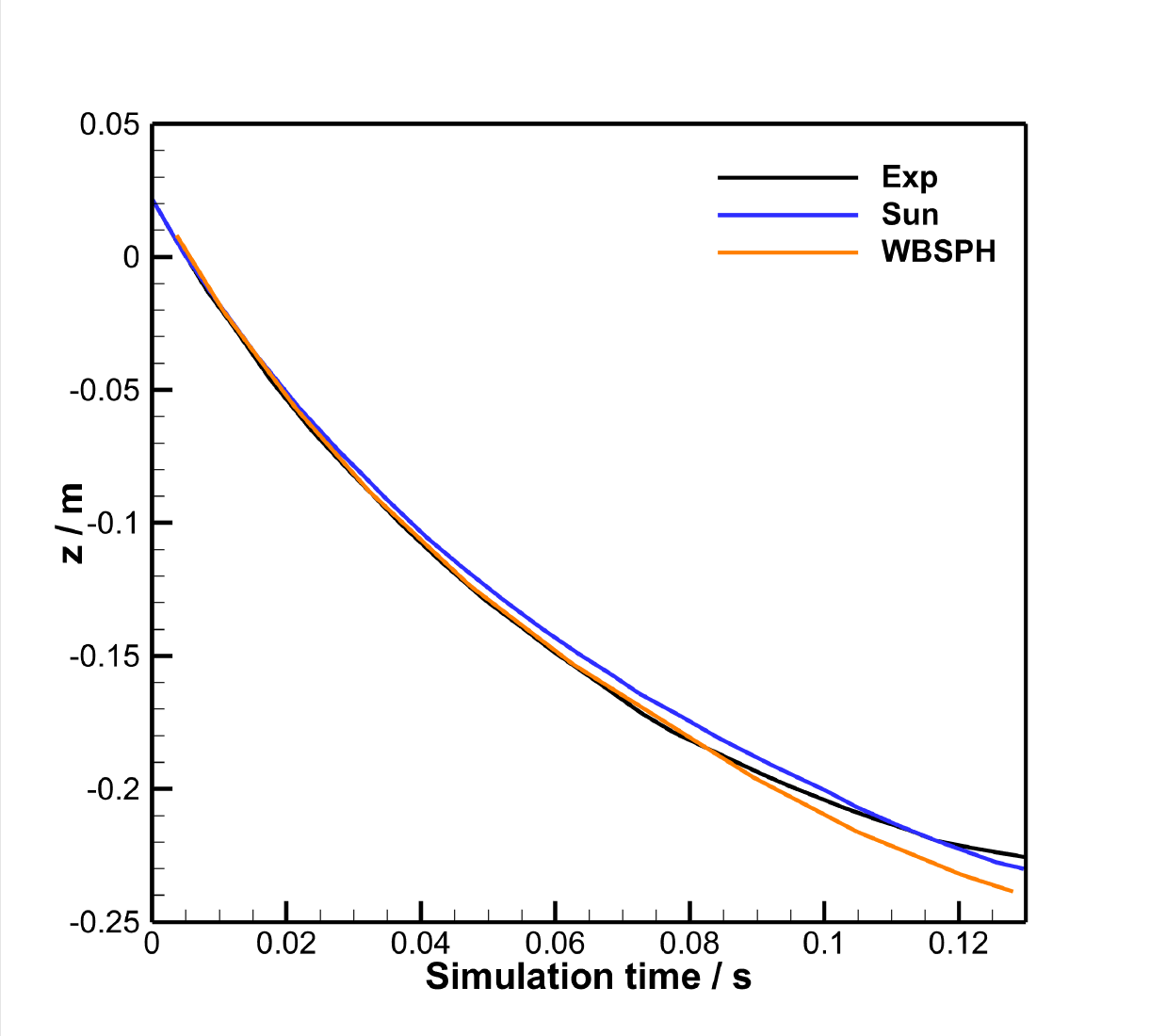}
    \caption{Cylinder entry problem: The trajectory of the horizontal cylinder by numerical simulations and experiment data.}
    \label{fig:fig8}
\end{figure}

Figure \ref{fig:fig8} shows the time evolution of the penetration depth of the horizontal cylinder. The present result agrees well with the experimental measurements \cite{waterentry-exp} and with the numerical prediction of Sun et al. \cite{sun1}. As the cylinder penetrates the water, the hydrodynamic resistance grows and progressively decelerates its descent, which is reflected in the gradual flattening of the trajectory. The close agreement of the predicted trajectory with the reference data indicates that the hydrodynamic force acting on the cylinder is accurately captured by the present method.

\begin{figure}[!htbp]
	\centering
	\subfigure[T=4ms]{\includegraphics[width = 0.2\columnwidth]{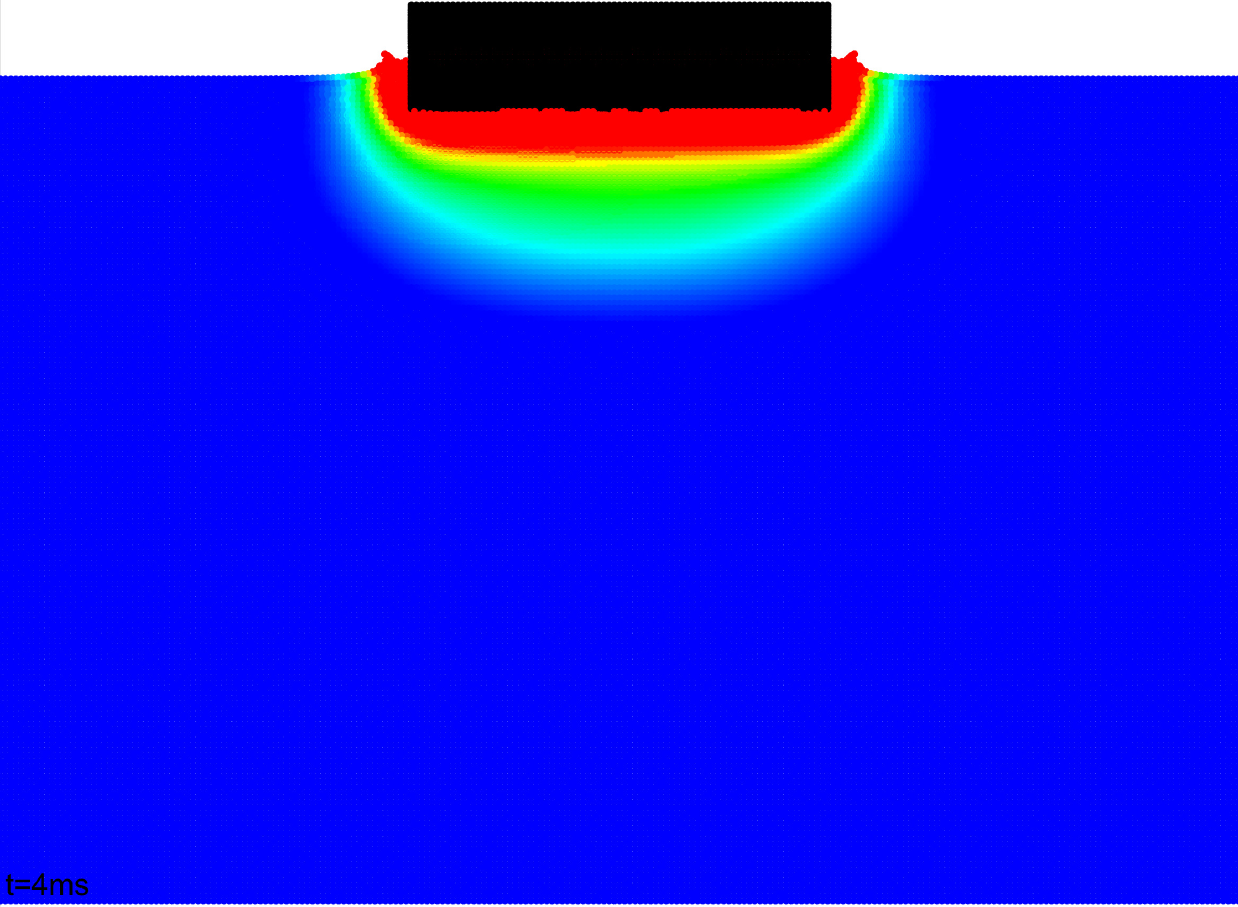}} \hspace{5pt}
	\subfigure[T=12ms]{\includegraphics[width = 0.2\columnwidth]{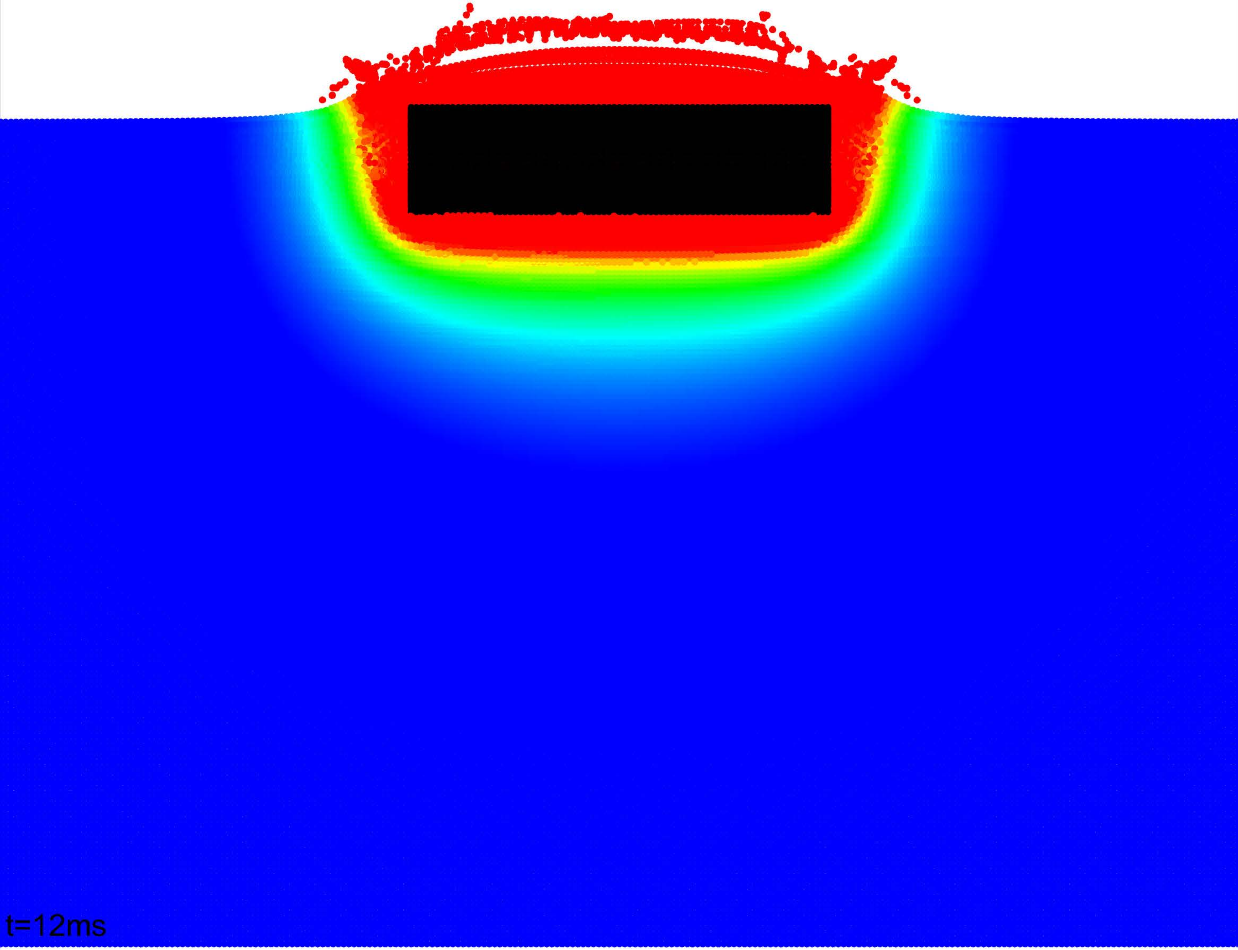}} \hspace{5pt}
    \subfigure[T=36ms]{\includegraphics[width = 0.2\columnwidth]{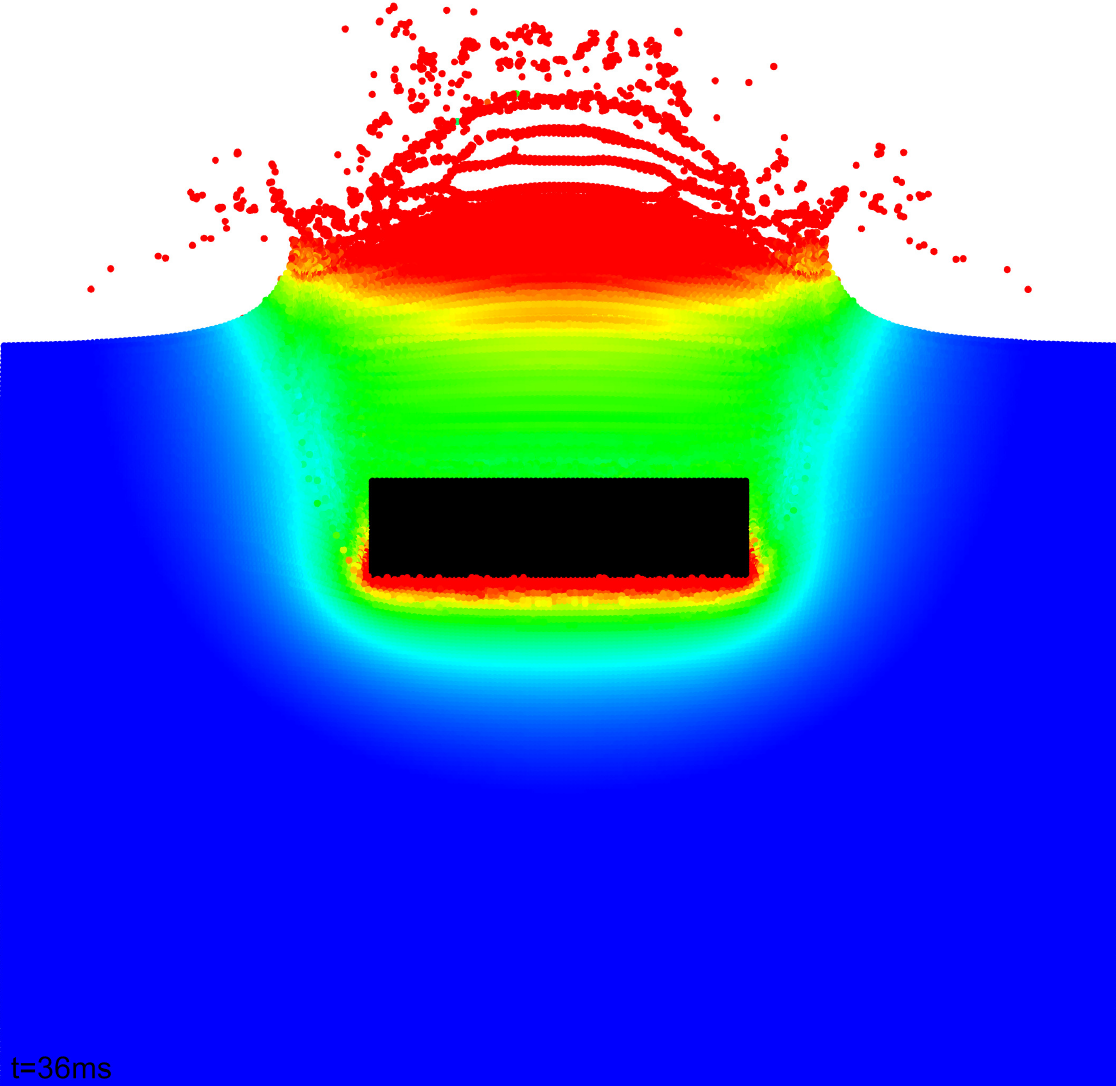}} \hspace{5pt}
    \subfigure[T=102ms]{\includegraphics[width = 0.2\columnwidth]{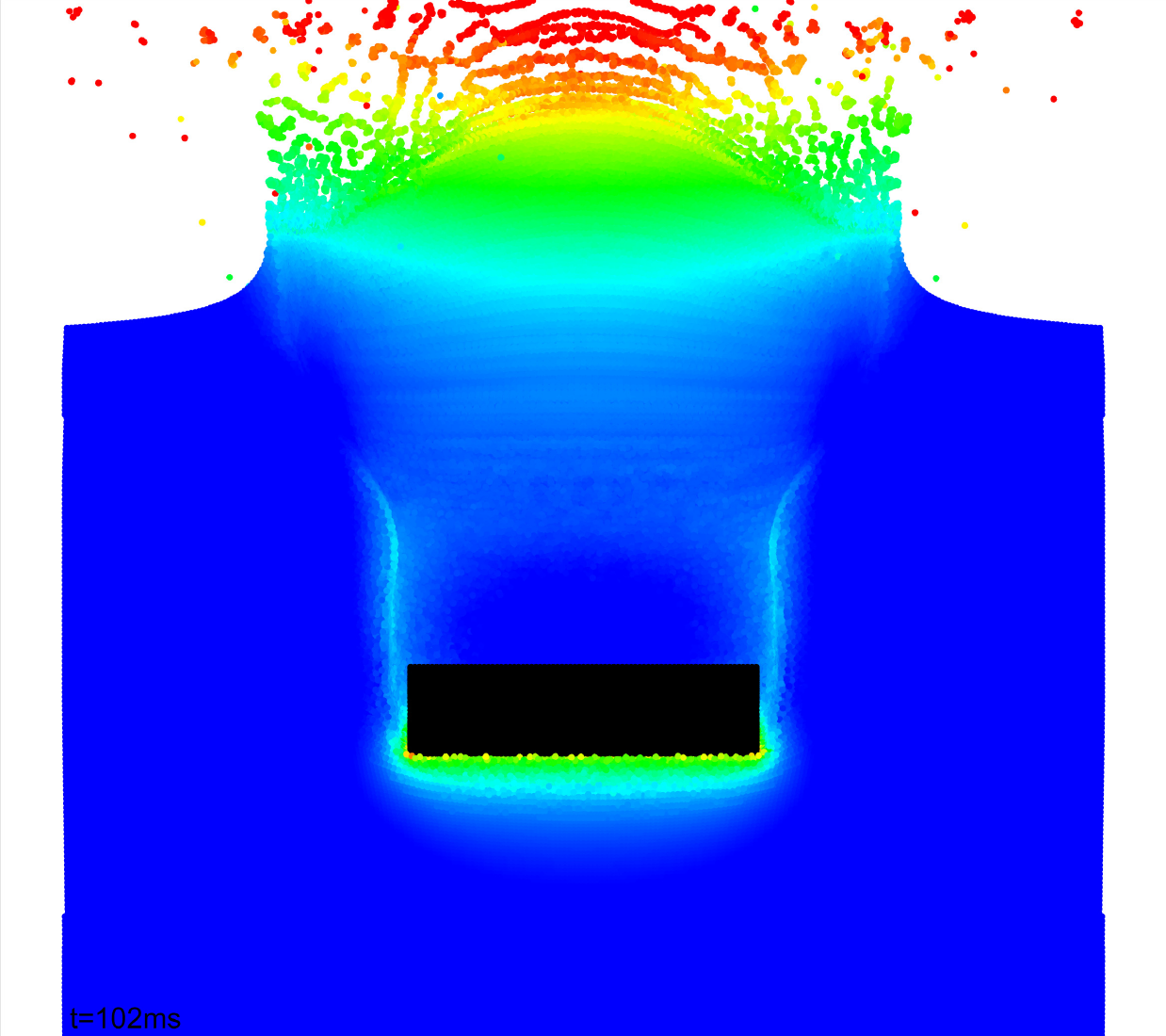}} \\
	\subfigure[T=120ms]{\includegraphics[width = 0.3\columnwidth]{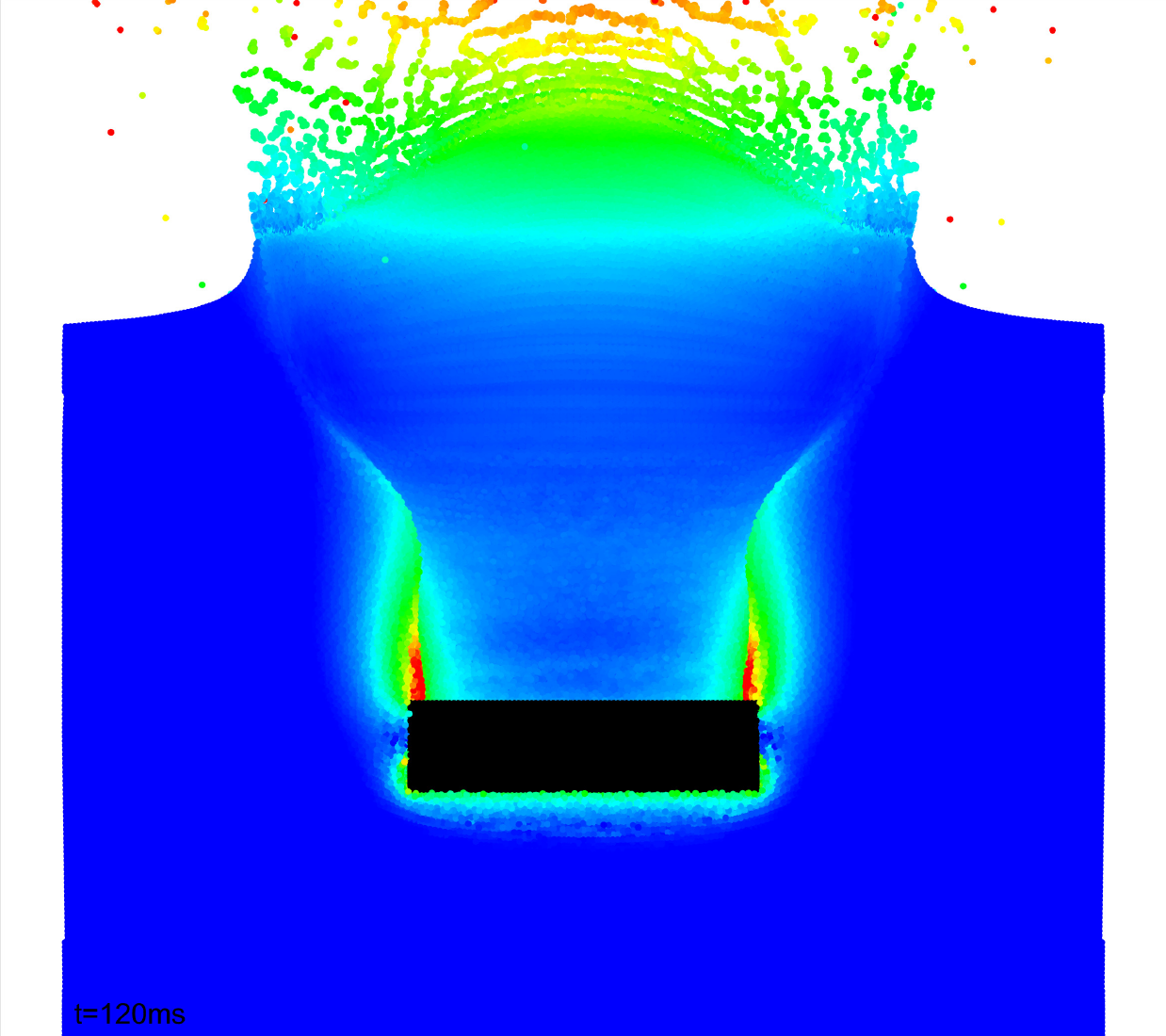}} \hspace{5pt}
    \subfigure[T=136ms]{\includegraphics[width = 0.3\columnwidth]{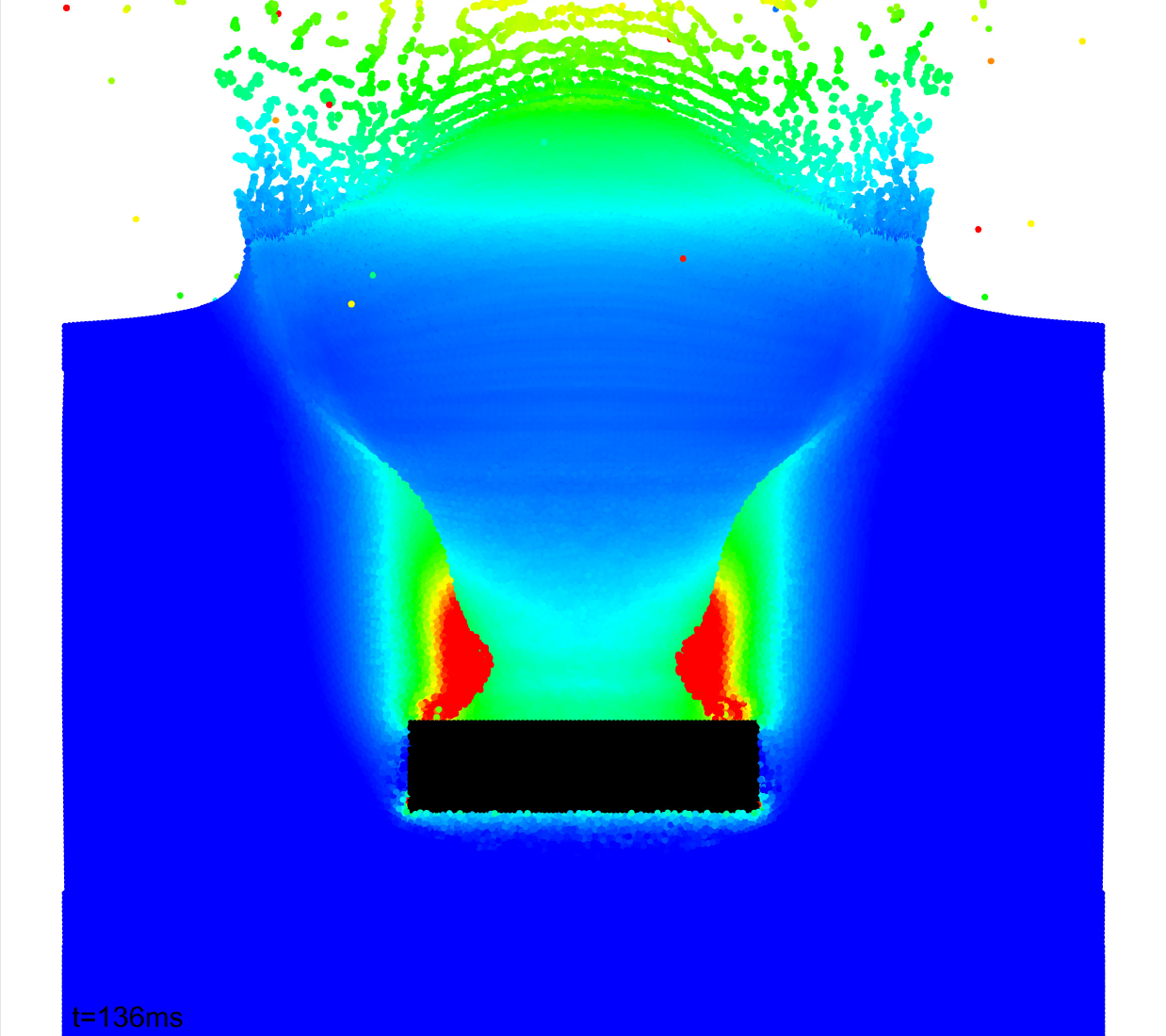}} \hspace{5pt}
    \subfigure[T=144ms]{\includegraphics[width = 0.3\columnwidth]{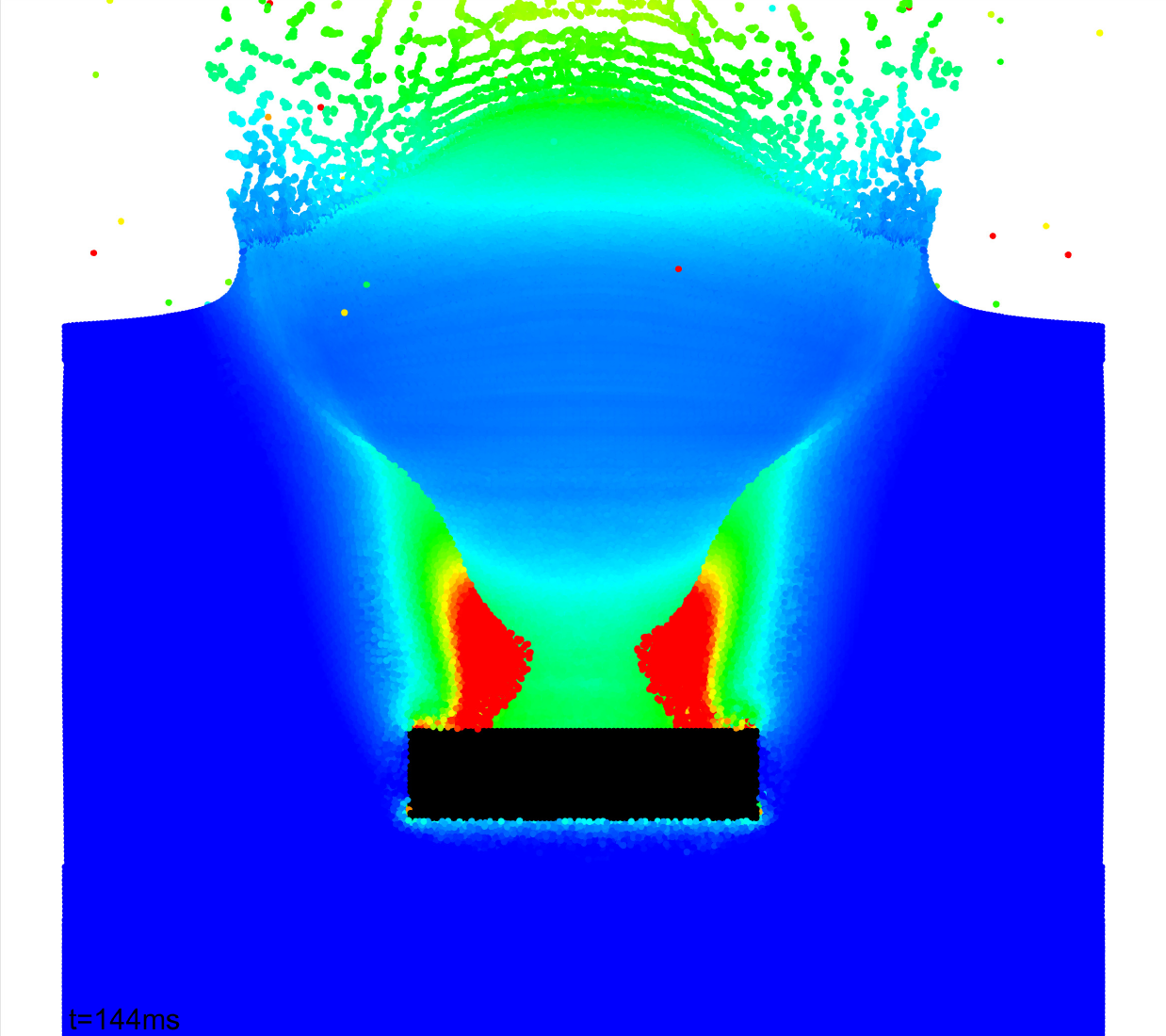}} \\
    \subfigure[Experimental sequences]{\includegraphics[width = \columnwidth]{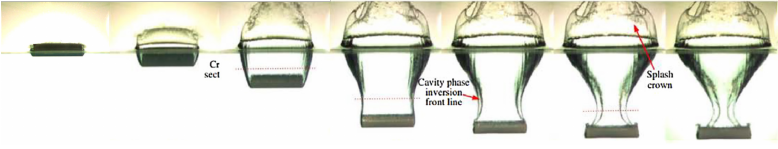}} \\
	\caption{Cylinder entry problem: Snapshots of horizontal cylinder colored by velocity magnitude at T = 4, 12, 36, 102, 120, 136, $144ms$. The initial impact velocity is $V_0 = -6.22m/s$. Top two rows: By WBSPH, bottom row: Experimental sequences.}
	\label{fig:fig9}
\end{figure}

Figure \ref{fig:fig9} presents a sequence of snapshots of the cylinder penetration and the accompanying free-surface deformation, colored by the velocity magnitude, at $T = 4$, $12$, $36$, $102$, $120$, $136$, and $144ms$. The successive stages of the process—the initial high-speed impact, the lateral discharge of water and the ensuing splash, the formation and expansion of the air cavity behind the cylinder, and its eventual pinch-off—are all reproduced by the present WBSPH, and the predicted free-surface morphology, together with the crown and cavity shapes, agrees well with the experimental sequence shown in the bottom row. The velocity fields remain smooth and free of spurious oscillations throughout this highly violent and strongly fragmented flow, further confirming the robustness and accuracy of the proposed formulation.

\subsection{Aircraft ditching problem}
\begin{figure}[!htbp]
	\centering
	\subfigure[Side view]{\includegraphics[width = 0.48\columnwidth]{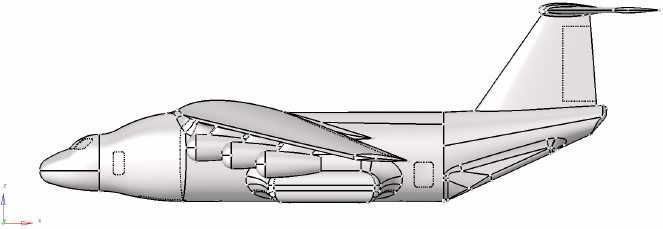}} \hspace{5pt}
	\subfigure[Front view]{\includegraphics[width = 0.48\columnwidth]{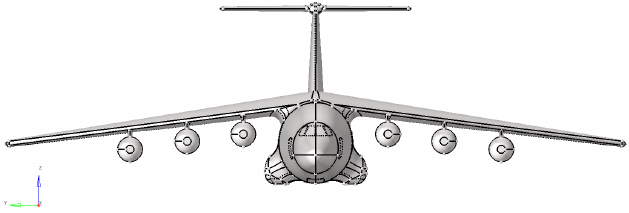}} 
	\caption{Aircraft ditching problem: Geometry of the model aircraft: (a) Side view, (b) Front view.}
	\label{fig:fig10}
\end{figure}

In this case, a real-scale aircraft ditching at the sea level is tested to verify the robustness and capability of complex geometry of WBSPH. The model aircraft with the length $L = 5m$ and width $D = 6m$ lands in a reservoir of size $20 \times 8 \times 2m$. The geometry of the model aircraft is shown in Figure \ref{fig:fig10} with the front and side views, where the details, including wings, fuselage and engine compartments, are all considered in this case. 

\begin{figure}[!htbp]
    \centering
    \includegraphics[width=0.7\linewidth]{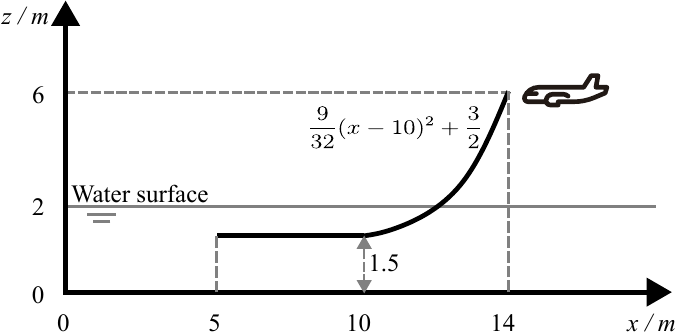}
    \caption{Aircraft ditching problem: Trajectory of the model aircraft.}
    \label{fig:fig11}
\end{figure}

The initial particle size is about $\Delta x = 0.03m$, which gives a total of about 11 millions particles. In this case, the model aircraft moves at a prescribed trajectory, the schematic trajectory of the model aircraft is shown in Figure \ref{fig:fig11}, where the aircraft impacts the water surface and glides to stop underwater. The time evolution of U-component velocity of the model aircraft is described by
\begin{equation}
    U = 
    \left \{
\begin{aligned}
    -5, \qquad &T \le 0.8s, \\
    \frac{5}{2}(T-3), \qquad &0.8s < T \le 2.8s.
\end{aligned}
    \right.
    \nonumber
\end{equation}

\begin{figure}[!htbp]
	\centering
    \setcounter{subfigure}{0}
	\subfigure[T=0.4]{\includegraphics[width = 0.48\columnwidth]{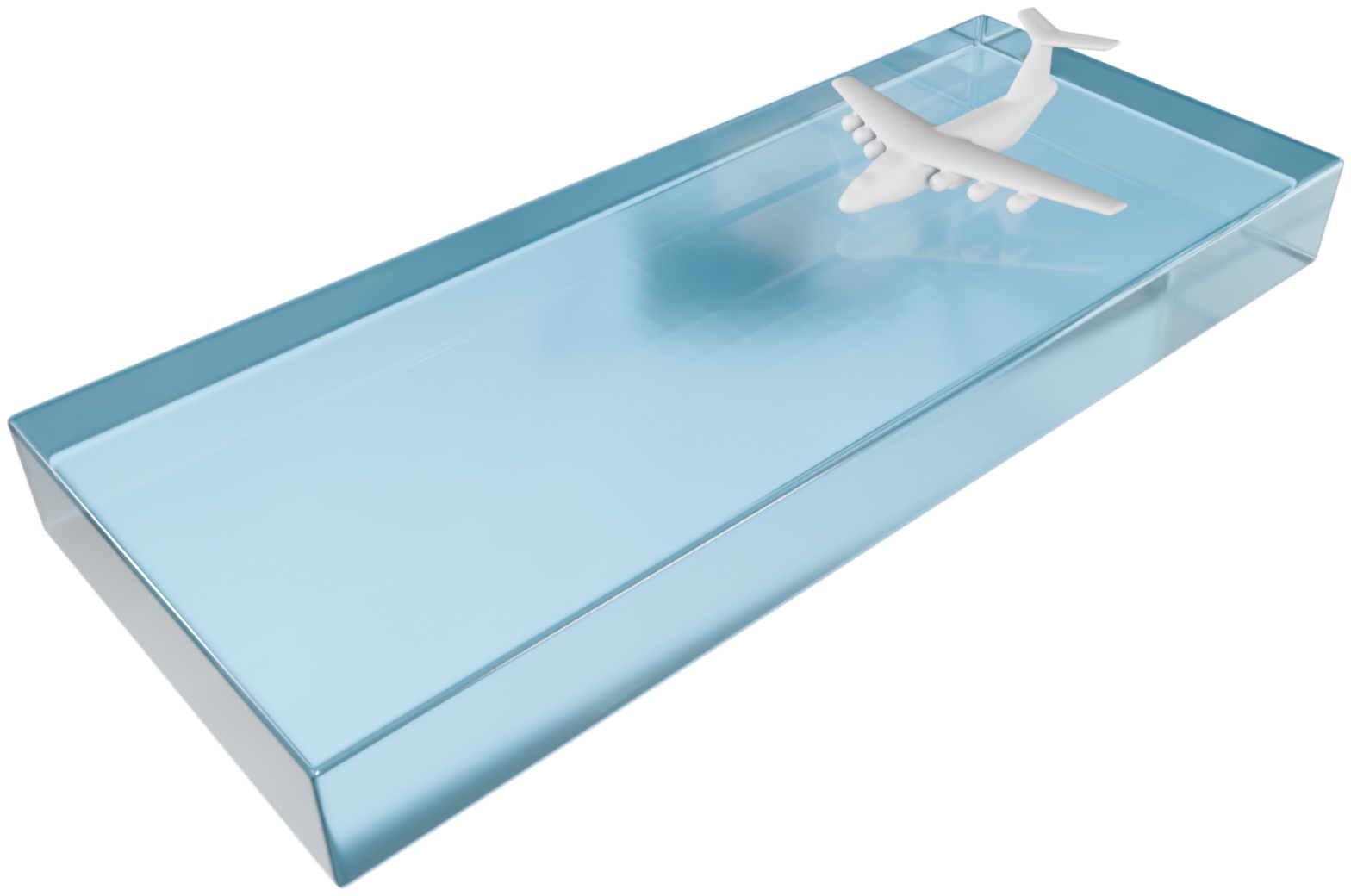}} \hspace{5pt}
	\subfigure[T=1]{\includegraphics[width = 0.48\columnwidth]{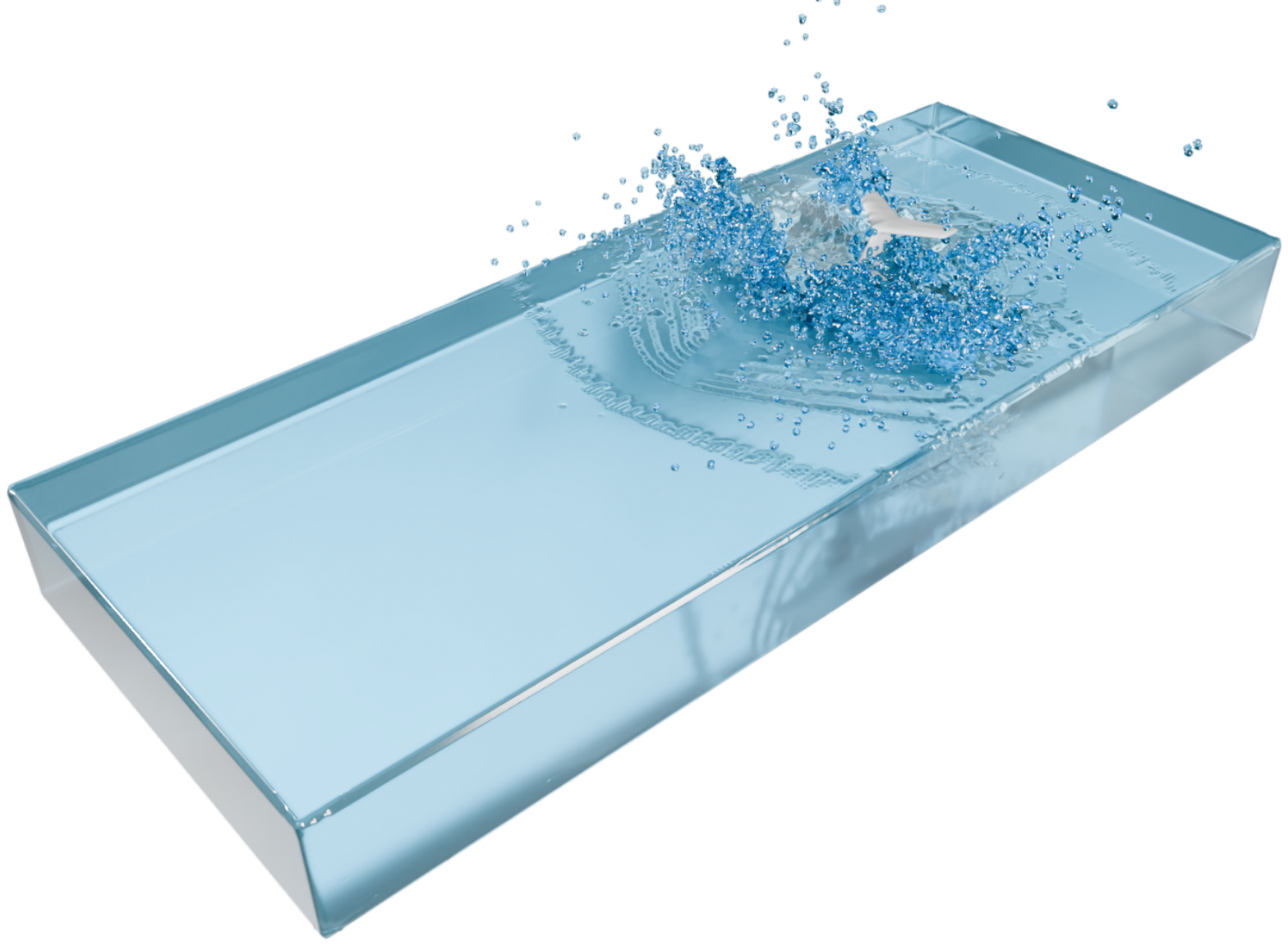}} \\
    \subfigure[T=1.4]{\includegraphics[width = 0.48\columnwidth]{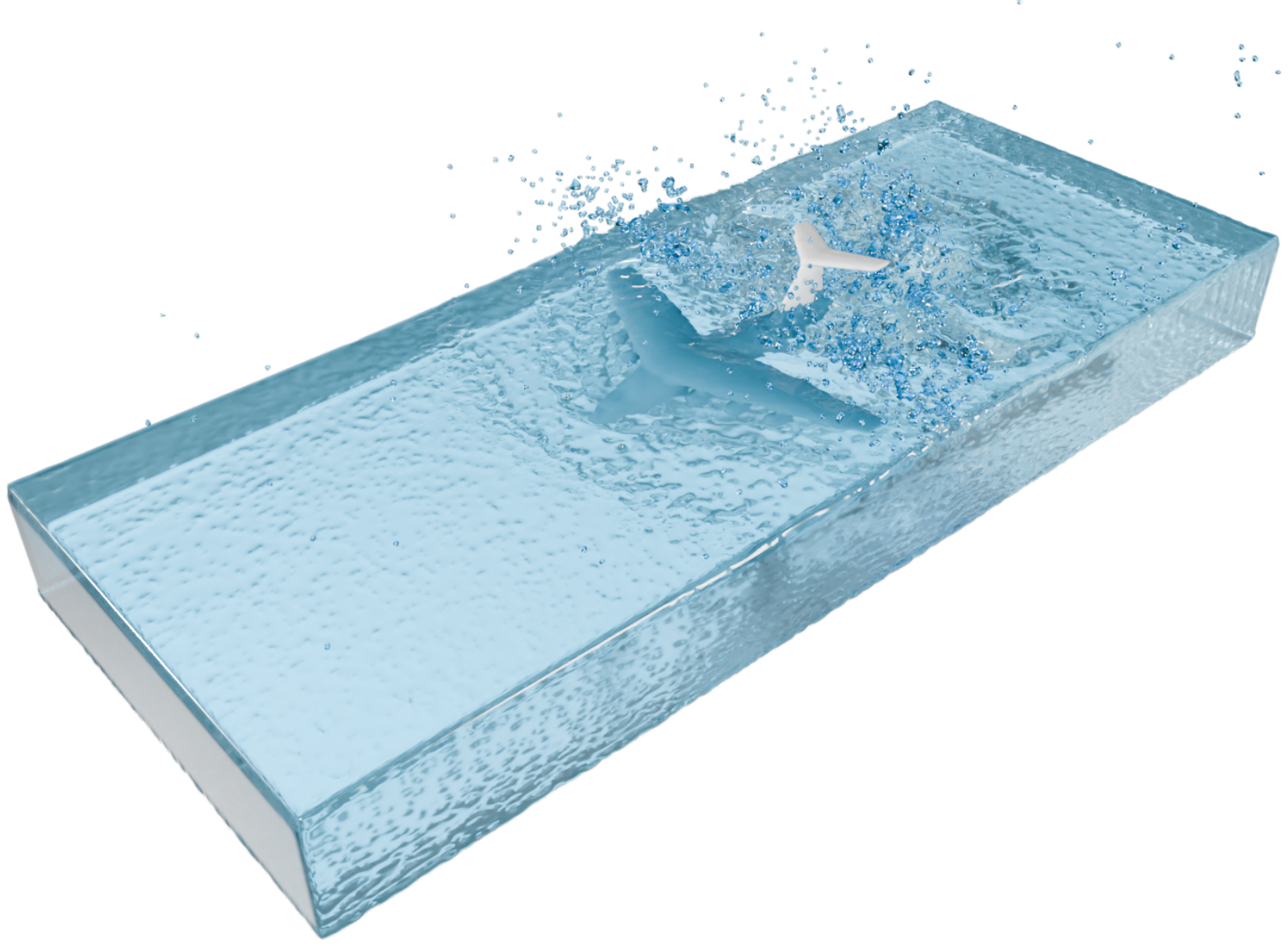}} \hspace{5pt}
	\subfigure[T=2.8]{\includegraphics[width = 0.48\columnwidth]{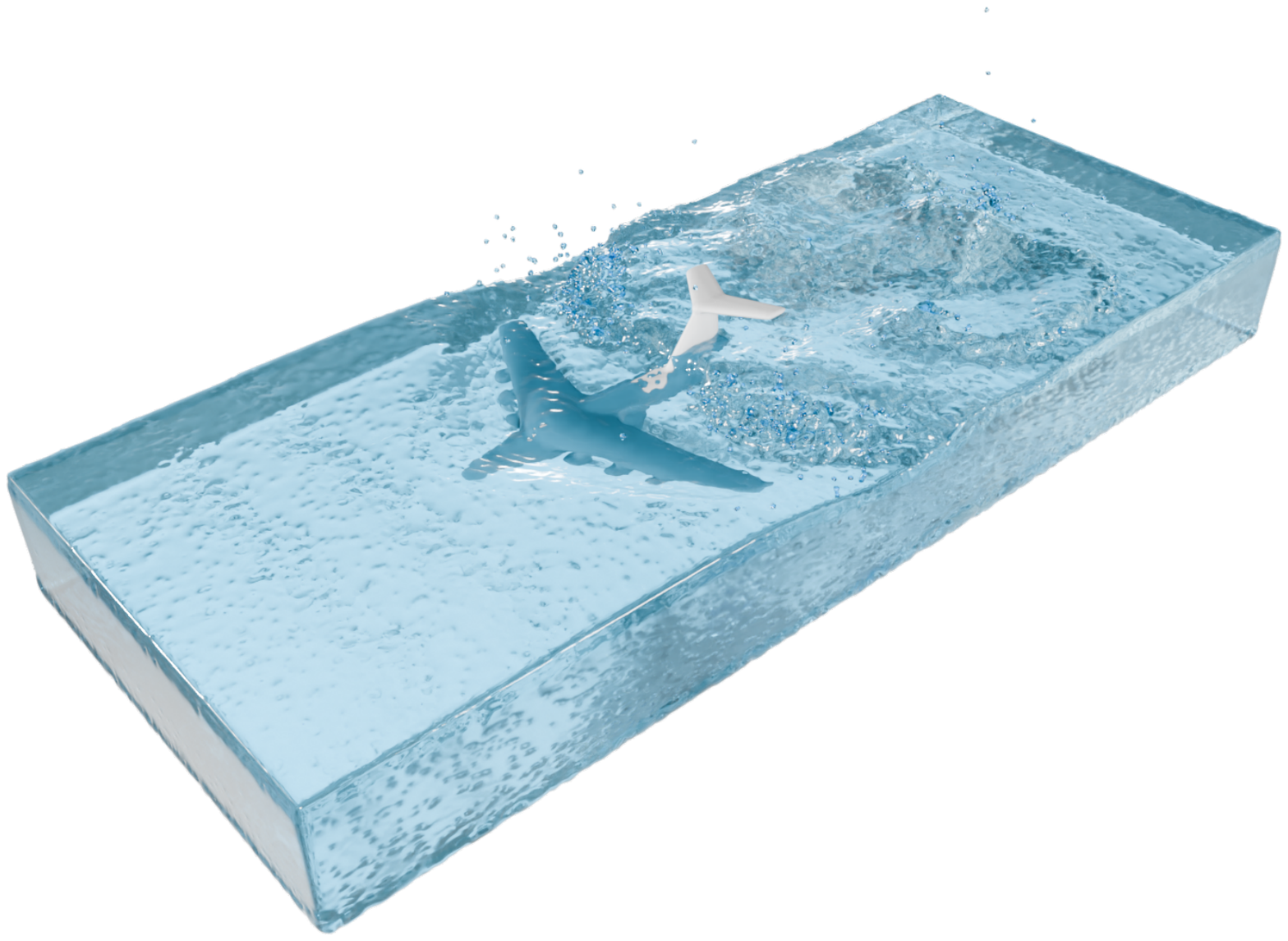}} \\
	\caption{Aircraft ditching problem: Time evolutions of the full-scale aerocraft ditching process at $T = 0.4$, $1$, $1.4$, and $2.8s$, from high-speed impact at the water level to substantially sliding forward.}
	\label{fig:fig12}
\end{figure}

Figure \ref{fig:fig12} shows the time evolution of the full-scale aircraft ditching process at $T = 0.4$, $1$, $1.4$, and $2.8s$. The aircraft descends towards the initially quiescent water surface before $T=0.4s$, and impacts the water surface after then. At $T = 1s$, the aircraft strikes the water and the high-speed impact generates an intense spray of fragmented sheets and droplets around the impact region. At $T = 1.4s$, the aircraft has penetrated further into the water and is gliding forward, with the wings ploughing through the free surface and producing two elongated splash sheets on either side, the free surface becomes strongly aerated and highly disturbed. By $T = 2.8s$, the aircraft has decelerated substantially and settles into the water while sliding forward, leaving behind an extensive wake of fragmented water and residual waves that gradually decay. Throughout the entire ditching process, the proposed WBSPH formulation captures the violent impact, the massive splash and fragmentation, and the subsequent deceleration and settling of the aircraft in a stable manner, demonstrating its applicability to large-scale, real-world engineering problems.

\subsection{Computational time analysis}
In this section, we compare the computational time of the three-dimensional dam breaking, cylinder entry and airplane ditching problems for 1000 steps on single CPU core (Intel Xeon CPU E5-2698 v3 @ 2.30GHz) with that on single NVIDIA RTX 3090 GPU. 

Table \ref{tab:efficiency} compares the computational time of the present solver on a single CPU core and on a single GPU for the three-dimensional dam break against an obstacle, the cylinder water entry, and the aircraft ditching cases. The total particle numbers, including both dummy (boundary) and real (fluid) particles, range from about 7.59 M for the dam break to 17.53 M for the cylinder entry, which is representative of the resolution required by realistic engineering simulations. The speedup is thus consistently of the order of $3000\times$ and shows little sensitivity to the specific configuration, indicating that the present GPU implementation maintains a high parallel efficiency across different geometries and particle numbers and shows the possibility to run large-scale industrial SPH simulations on personal computers.

\begin{table}[!htbp]
		\small
		\begin{center}
			\def\temptablewidth{1.0\textwidth}
			{\rule{\temptablewidth}{1pt}}
			\begin{tabular*}{\temptablewidth}{@{\extracolsep{\fill}}c c c c c}
				Case & Number of particles (dummy + real) & CPU / s & GPU / s & speedup ratio\\
				\hline
                Dam breaking & 3.29M + 4.30M = 7.59M & $2.28\times10^{5}$ & 79.40 & 2869 \\
                Cylinder entry & 2.37M + 15.16M = 17.53M & $8.53\times 10^{5}$ & 270.31 & 3153 \\
                Ditching & 3.60M + 11.85M = 15.45M & $5.53\times 10^{5}$ & 179.00 & 3089 \\
			\end{tabular*}
			{\rule{\temptablewidth}{1pt}}
		\end{center}
		\vspace{-4mm} \caption{Comparison of the computational time of WBSPH on a single CPU core and a single GPU device for the aforementioned cases.}
		\label{tab:efficiency}
\end{table}

\section{Conclusions}
This study has proposed a well-balanced weakly compressible SPH formulation for the Navier–Stokes equations that preserves the analytical hydrostatic equilibrium solution exactly at the discrete level, regardless of the bottom topography, and has extended the formulation to three dimensions with an efficient GPU implementation. The exact discrete balance is achieved through two complementary ingredients. First, the Navier–Stokes equations are recast into a new form through the introduction of an auxiliary potential variable, which converts the nonlinear pressure-gradient-over-density term into the gradient of a single scalar field that reduces to a linear function of position under hydrostatic conditions. Second, a Riemann-based particle approximation of the potential gradient with first-order consistency is developed to balance the constant gravitational force exactly. The stabilization techniques widely used in conventional weakly compressible SPH—including $\delta-$SPH, particle shifting and tensile instability control techniques—are readily incorporated into the new framework, so that the well-balanced property is obtained without sacrificing the robustness required for violent free-surface flows. 

Hydrostatic test cases with rectangular, triangular, and Gaussian bottom topographies show that the proposed formulation reduces the spurious velocity error from the $10^{-2}-10^{-3}$ level of conventional SPH to the order of $10^{-14}-10^{-13}$, essentially at machine precision, while preserving the free-surface elevation exactly over long-term evolution. A series of three-dimensional benchmark problems, including the dam breaking problem against an obstacle, the water entry of a horizontal cylinder, and a full-scale aircraft ditching problem, demonstrate the robustness, accuracy and low pressure oscillation of the well-balanced SPH formulation under violent impact, splashing, and free-surface fragmentation. Computational-time analyses over the three-dimensional problems demonstrate that the GPU implementation attains a speedup of up to $3000\times$ on a single consumer-grade NVIDIA RTX 3090 GPU relative to a single CPU core (Intel Xeon E5-2698 v3 @ 2.30 GHz), enabling simulations with up to 17.53 million particles on a personal computer. Future work will focus on extending the framework to multiphase flows and multi-GPU implementation for even larger-scale applications. Overall, the present methodology provides a solid foundation for the development of well-balanced SPH formulations for long-term simulations of free-surface flows.

\section*{Acknowledgements}
This research was supported by the Research Grants Council Areas of Excellence (AoE) Scheme (AoE/P-601/23N-D-MATH), and by CORE as a joint research center for ocean
research between Laoshan Laboratory and HKUST.

\bibliographystyle{ieeetr}
\bibliography{reference}

\end{document}